\documentclass[aip,jcp,amsmath,amssymb,preprint,superscriptaddress, footinbib]{revtex4-1}

\usepackage{rgRPA_parameters}

\usepackage{xr-hyper}

\usepackage{xcolor}
\usepackage{cancel}

\usepackage{graphicx}

\begin{document}

\title
{\maintitle}

\author{Yi-Hsuan Lin}
\affiliation
{Department of Biochemistry, University of Toronto, Toronto, Ontario, Canada}
\affiliation
{Molecular Medicine, The Hospital for Sick Children, Toronto, Ontario, Canada}
\author{Jacob P. Brady}
\affiliation
{Department of Molecular Genetics, University of Toronto, Toronto, Ontario, Canada}
\affiliation
{Department of Chemistry, University of Toronto, Toronto, ON, Canada}
\affiliation
{Department of Biochemistry, University of Toronto, Toronto, Ontario, Canada}
\author{{\hbox{Hue Sun Chan}}}
\email{chan@arrhenius.med.utoronto.ca}
\affiliation
{Department of Biochemistry, University of Toronto, Toronto, Ontario, Canada}
\author{Kingshuk Ghosh}
\email{kingshuk.ghosh@du.edu}
\affiliation
{Department of Physics and Astronomy, University of Denver, Colorado, CO, USA}
\affiliation
{Molecular and Cellular Biophysics, University of Denver, Colorado, CO, USA}

\date{\today}

\begin{abstract}
The physical chemistry of liquid-liquid phase separation (LLPS) of polymer 
solutions bears directly on the assembly of biologically functional 
droplet-like bodies from proteins and nucleic acids. These biomolecular 
condensates include certain extracellular materials, and intracellular 
compartments that are characterized as ``membraneless organelles''.  
Analytical theories are a valuable, computationally efficient tool for 
addressing general principles. LLPS of neutral homopolymers 
are quite well described by theory; but it has been a challenge to develop
general theories for the LLPS of heteropolymers involving 
charge-charge interactions. Here we present a theory that combines a 
random-phase-approximation treatment of polymer density fluctuations 
and an account of intrachain conformational heterogeneity based upon 
renormalized Kuhn lengths to provide predictions of LLPS properties as 
a function of pH, salt, and charge patterning along the chain sequence.  
Advancing beyond more limited analytical 
approaches, our LLPS theory is applicable to a wide variety
of charged sequences ranging from highly charged polyelectrolytes to 
neutral or nearly neutral polyampholytes. This theory 
should be useful in high-throughput screening of protein and
other sequences for their LLPS propensities and can serve as a basis 
for more comprehensive theories that incorporate
non-electrostatic interactions. Experimental ramifications of our
theory are discussed. 

\end{abstract}

\maketitle

\section{Introduction}
Mesoscopic compartmentalization undergirded by 
liquid-liquid phase separation (LLPS) of intrinsically disordered proteins
or regions (IDPs or IDRs) and nucleic acids is now recognized as a 
versatile means for biomolecular organization and 
regulation~\cite{Cliff2009,Rosen2012,McKnight2012,Nott2015,MolliexCell2015,PakMolCell2016}.
Some of these phase-separated droplet-like compartments are intracellular 
bodies---such as stress granules, P-granules and nucleoli---that may
be characterized
as ``membraneless organelles''. Outside the cell, biomolecular 
LLPS can be biologically useful 
as well, as in the formation of certain extracellular 
materials. Collectively referred to as biomolecular condensates, these 
phase-separated bodies participate in many vital functions, as highlighted
by their recently elucidated roles in
endocytosis~\cite{BergeronArxiv2018}, 
silencing chromatin~\cite{LarsonNature2017}, 
transcription~\cite{PlysKingstonScience2018,ChoScience2018,SabariScience2018},
and translation~\cite{BrianTsang2019}.
The repertoire of relevant discoveries is rapidly 
expanding~\cite{Cliff2017Rev,Rosen2017Rev,Monika2018Rev}. LLPS of globular proteins, for 
example lens protein solutions, have also been observed and are of biological importance 
\cite{BenedekPNAS1991,BenedeckPRL1996,PalmaProteins1999,ZhouPang,ZhouJPCB2016,ChanWinterJACS}.

Recent bioinformatics analyses suggest that IDPs and IDRs comprise a 
significant fraction of the proteomes of higher organisms, and that 
functional LLPS is likely ubiquitous~\cite{JDFKJMB2018}. 
The propensity for an IDP or IDR to phase separate is governed by its 
amino acid sequence and modulated by solution/environmental conditions 
(temperature, hydrostatic pressure~\cite{Roland2019Rev}, pH, 
ionic strength~\cite{Brady2017,Alberti2017}, etc) as well as their 
interactions with other biopolymers such as RNA.
Thus, any ``big-picture'' survey of the physical basis of
biomolecular condensates requires not only consideration of many different 
sequences but a large variety of environmental conditions.  Adding 
to this combinatorial complexity is that even for a given wildtype sequence, 
postranslational modifications, mutations, and 
splicing~\cite{Nott2015,MonahanEMBO2017}
can lead to diverse LLPS propensities. In this context,
analytical theories are the most computationally efficient tool for large-scale
exploration of sequence-dependent biomolecular LLPS. Although explicit-chain
simulations provide more energetic and structural details~\cite{MittalBestPlosCompBio2018,Das2018b,DignonPNAS2018} and field-theory simulations afford more 
numerical accuracy~\cite{McCartyJPCL2019,DanielsenPNAS2019,FredricksonJCP2019}, 
currently the number of sequences that can be simulated by these approaches is 
limited because of their high computational cost. Moreover, analytical
theories are valuable for insights into physical principles
that are less manifest in simulation studies.
With this in mind, we build on recent success in using analytical 
theories to account for sequence-dependent biomolecular condensates under
certain limited conditions~\cite{RohitNaturePhysics2015,Biochem2018} so as to 
develop improved theories that are 
more generally applicable. 

Building sequence-specific theories of LLPS will also have implications in phase separation of block polyamphoytes and its comparison with complex coacervation between
oppositely charged homopolyelectrolytes, a topic of intense research in polymer physics \cite{JongKruyt1929,OverbeekVoorn1957,SpruijtMacro2010,TirrellMacro2010,PerryTirrell2014,PerryandSing2015,SrivastavaTirrell2016,Lytle2016, Lytle2017,SingMacro2017,CoacervationSoftMatterRev,Zhang2018,MuthuJCP2018,TirrellMacro2018,PerrySing2019}. 
Diblock polyampholytes with repeat units of a polycation segment followed by a polyanion segment can be envisioned to be equivalent to two oppositely charged homopolyelectrolytes. For this reason LLPS of block polyampholytes -- a limiting case of our theory -- is often termed self-coacervation \cite{FredricksonJCP2019,DanielsenPNAS2019}
and shares features similar to complex cocervation of a polycation and polyanion \cite{PerrySing2019}. Experiments and simulation have also reported differences between
the phase diagrams of block polyampholytes and homopolyelectrolyte coacervation. The observed differences can be explained by the presence of `charge pattern interfaces' where two segments of oppositely charged blocks merge in polyampholytes. Homopolyelectrolytes, on the other hand, lack such connectivities, thus leading to different types of salt localization in comparison to block polyampholytes \cite{PerrySing2019}. Application of a general sequence-based analytical theory of polyampholyte LLPS will further advance these comparisons between complex coacervation and self-coacervation. Future effort in theory development is needed in this direction. Thus, our framework should be useful not only for high-throughput analyses of the LLPS propensities of naturally occurring biological sequences but also for the design of artificial biological and non-biological heteropolymers with desired LLPS properties~\cite{SingPerryNatComm2017,Chilkoti2018,PerrySingACSCentSci2019}.


Inasmuch as sequence-specific analytical theories for biomolecular condensates 
are concerned, a recent multiple-chain formulation based on the traditional
random phase approximation (RPA)~\cite{Mahdi2000,Ermoshkin2003} has been 
applied to study the dependence of LLPS of IDPs on the charge patterns along 
their chain sequences~\cite{Lin2016}.
This approach accounts for the experimental difference 
in LLPS propensity between the Ddx4 helicase IDR and its charge-scrambled
mutant~\cite{Lin2016,Lin2017a}. It also provides insight into a possible
anti-correlation between multiple-chain LLPS propensity and single-chain 
conformational dimensions~\cite{Lin2017b} as well as the degree of 
demixing of different charge sequences under LLPS conditions~\cite{Lin2017c}.
As an initial step, these advances are useful. As a heteropolymer 
theory, however, traditional RPA~\cite{Mahdi2000,Ermoshkin2003} is known
to have two main shortcomings. First, the density of monomers of the polymer
chains in solution is assumed to be roughly homogeneous 
as density fluctuations are neglected beyond second order in RPA. A rigorous 
treatment proposed by Edwards and Muthukumar has shown the importance of 
including density fluctuations to higher 
orders~\cite{Muthu1996,MuthuEdwards2016,MuthuMacro2017}. 
Nonetheless, a recent comparison of field-theory simulation and RPA indicates
that RPA is reasonably accurate for intermediate to high monomer densities 
for the cases considered, and that significant deviations between RPA and 
field theory simulation occur only for volume fraction 
$< 0.02$ that of the highest condensed-phase simulated~\cite{McCartyJPCL2019}.
Second, traditional RPA neglects the fact that monomer-monomer interactions
can cause conformational variation of individual chains by computing
the single-chain structure factor using a Gaussian chain with no 
intrachain interaction. This limitation, which applies to homopolymers as
well as heteropolymers, is particularly acute for the latter.
Indeed, experimental and computational studies have shown
that single-chain conformational heterogeneities and dimensions
are sensitive to sequence specific
interactions~\cite{HofmannPNAS2012,DasPappu2013,SchulerAnnRevBiophys2016,KonigNatMethods2015,SorannoPNAS2014,VaianaBJ2015}. 
Regarding this shortcoming, recently an improved analytical approach was 
developed at the single-chain level by replacing the Kuhn length $l$ (termed 
``bare'' Kuhn length) of the Gaussian chain by a set of renormalized Kuhn 
lengths, $l_1$, that embodies the sequence-specific interactions 
approximately~\cite{Sawle2015,Firman2018,HuihuiFirmanGhoshJCP2018}. 
Renormalized structure factors have also been exploited to improve
homopolymer LLPS theories for polyelectrolytes~\cite{Shen2017,Shen2018}.

Noting that the first shortcoming described above is likely limited only to 
regimes of extremely low polymer concentrations, here we first focus 
on rectifying the second shortcoming by combining the earlier, traditional 
sequence-dependent RPA theory~\cite{Lin2016,Lin2017a} with the 
sequence-dependent single-chain theory that utilizes a renormalized Gaussian 
(rG) chain formulation~\cite{Sawle2015,Firman2018,HuihuiFirmanGhoshJCP2018}
for a better account of conformational heterogeneity. We refer to this
theory as rG-RPA. As a control, we also study a simpler theory, 
analogous to our earlier formulation~\cite{Lin2016,Lin2017a}, 
that invokes a Gaussian chain with fixed Kuhn length. Following 
Shen and Wang~\cite{Shen2018}, we refer to this $l_1=l$ theory as fG-RPA. 
Extensive comparisons of rG-RPA and fG-RPA predictions 
on various systems indicate that rG-RPA represents a significant improvement
over fG-RPA.  As will be detailed below,
the superiority of rG-RPA is most notable in its ability 
to account for the LLPSs of both polyampholytes and polyelectrolytes 
whereas fG-RPA is inadequate for polyelectrolytic polymers.


\section{Theory}

We consider an overall neutral solution of $\np$ charged polymers, each
consisting of $N$ monomers (residues), and
small ions including $\ns$ salt ions and $\nc$ counterions with
charge numbers $\zs$ and $\zc$ respectively.
The charge pattern of a polymer is given by
an $N$-dimensional vector
$\ket{\sigma} = [\sigma_1, \sigma_2, ..., \sigma_N]^{\rm T}$, where 
$\sigma_\tau$ is the charge on the $\tau$th monomer; and
$\pc \equiv (\sum_\tau \sigma_\tau)/N$ is the net charge per monomer. 
For simplicity, we consider the case with only one species of positive and one 
species of negative ions; their numbers are denoted as $\nsp$ and $\nsm$
respectively. Moreover, ``salt" is identified as 
the small ions that carry charges of the same sign 
as the polymers, whereas ``counterions" are the small ions 
carrying charges opposite to that of the polymers. 
Thus, $\ns=\nsp$ if $\pc >0$ and $\ns=\nsm$ if $\pc < 0$; and
$|\pc| \np N + \zs \ns = \zc \nc$ for solution neutrality.
The densities ($\rho$) of monomers, salt ions, and counterions are,
respectively, $\rho_m = \np N/\Omega$, $\rho_s = \ns/\Omega$, 
and $\rho_c = \nc/\Omega$, where $\Omega$ is solution volume.
Although only a simple system with at most two species of small ions is 
analyzed here for conceptual clarity, our theory can be readily expanded 
to account for multiple species of small ions.

Details of our formulation are given in the Appendix.
Here we provide the key steps in the derivation. Let
$F$ be the total free energy of the system. Then
$f \equiv F l^3/(\kB T\Omega)$ is free energy in units of $\kB T$ 
per volume $l^3$, where $l$ is the bare Kuhn length, $\kB$ is Boltzmann 
constant and $T$ is absolute temperature. In our theory,
\begin{equation}
f = -s + \fion + \fpoly + \fzero \; ,
\label{f_eq}
\end{equation}
where $s$ is mixing entropy, $\fion$ and $\fpoly$ are interactions 
among the small ions and involving the polymers, respectively, 
that arise from density fluctuations
and $\fzero$ is the mean-field excluded volume interaction,
all expressed in the same units as $f$. The mixing entropy, which accounts
for the configurational freedom of the solutes, takes the Flory-Huggins
form, viz.,
\begin{equation}
-s = \frac{\phi_m}{N}\ln\phi_m + \phi_s\ln\phi_s + \phi_c \ln \phi_c + 
\phi_w\ln\phi_w \; ,
\label{eq_mix}
\end{equation}
where $\phi_m$, $ \phi_s$, $ \phi_c$, and 
$\phi_w = 1\!-\!\phi_m\!-\!\phi_s\!-\!\phi_c$ are volume fractions
($\phi=\rho l^3$),
respectively, of polymers, salt ions, 
counterions, and solvent (water for IDP systems).
Following Muthukumar,
the charge of each small ion is taken to be distributed over a finite 
volume comparable to that of a monomer.
The corresponding interaction free energy among the small ions 
is~\cite{Muthu2002} 
\begin{equation}
\fion  = -\frac{1}{4\pi}\left[ \ln(1+\kappa l) - \kappa l + \frac{1}{2}(\kappa l)^2 \right] \; ,
\end{equation}
where $1/\kappa = 1/\sqrt{4\pi\lb(\zs^2\rho_s+\zc^2\rho_c)}$ is the Debye 
screening length, $\lb$ being Bejurrm length.
Polymers interact via a $\kappa$-dependent screened 
Coulomb potential and a uniform excluded-volume repulsion with 
strength $v_2$. The origin of this repulsive term is to be understood as an effective
interaction between polymer and solvent. By setting $v_2$ repulsive we imply the polymer
is in a good solvent.
These interactions are contained in the expression
\begin{equation}
\U_p[\R] = \frac{1}{2} \sum_{\alpha,\beta=1}^\np\sum_{\tau,\mu=1}^N 
\left[
\frac{\sigma_\tau\sigma_\mu e^{-\kappa|\R_{\alpha,\tau}-\R_{\beta,\mu}|}}{|\R_{\alpha,\tau}-\R_{\beta,\mu}|}
+ \wtwo\delta^3\left(\R_{\alpha,\tau}-\R_{\beta,\mu} \right)\right] \; ,
\end{equation} 
where $\R_{\alpha,\tau}$ is the position of the $\tau$th monomer 
in the $\alpha$th polymer. The $\U_p$ form facilitates the formulation
in terms of density fields below. For this purpose, the divergent
self-interaction terms in $\U_p$ are either regularized subsequently
or inconsequential because they 
do not contribute to phase-separation properties.
Chain connectivity of the polymers are enforced by the potential 
\begin{equation}
\T[\R] = \frac{3}{2l^2}\sum_{\alpha=1}^{\np}\sum_{\tau=1}^{N-1}
\left(\R_{\alpha,\tau+1}-\R_{\alpha,\tau}\right)^2 \; .
	\label{eq:Tini}
\end{equation} 
Thus, aside from a combinatorial factor that has already been included
in Eq.~\ref{eq_mix}, 
the partition function involving the polymers is given by 
\begin{equation}
\Zp = \int \prod_{\alpha=1}^{\np}\prod_{\tau=1}^N d\R_{\alpha,\tau}e^{-\T[\R]-\U_p[\R]} \; .
	\label{eq:Zini}
\end{equation}
Now, by applying the Hubbard-Stratonovich transformation and converting 
real-space to $\kk$-space variables, we convert the coordinate-space partition 
function in Eq.~\ref{eq:Zini} to a $\kk$-space partition 
function~\cite{McCartyJPCL2019, DanielsenPNAS2019} involving
a charge-density field $\psi$ and a matter-density field $\w$, viz., 
\begin{equation}
\Zp = \Zzero \Zp^\prime \; , \quad \Zp^\prime =  \int \prod_{\kk\neq\kzero} \sqrt{\frac{\vcol_k}{\wtwo}} 
\frac {d\psi_\kk d\w_\kk}{2\pi\Omega} e^{-\Eng[\psi, \w]} \; ,
	\label{eq:Zp_H}
\end{equation}
where $\Zzero=\exp[-\wtwo(N\np)^2/2\Omega]$ is the factor for $\kk={\bf 0}$, 
\begin{equation}
\Eng[\psi, \w] =  \frac{1}{2\Omega}\sum_{\kk\neq\kzero}\left[ 
				\vcol_k\psi_{-\kk}\psi_{\kk} 
				+ \frac{\w_{-\kk}\w_{\kk}}{\wtwo} \right]  - \np \ln \Qp[\psi,\w] \; ,
	\label{eq:Hall}
\end{equation}
$\vcol_k \equiv k^2/(4\pi\lb) + (\zs^2\rho_s + \zc^2\rho_c)$,
$k\equiv |\kk|$, 
$\Qp[\psi, \w] = \int\DD[\R] \exp(-\Hp[\psi,\w])$
is the single-polymer partition function with
$\DD[\R] \equiv \prod_{\tau=1}^Nd\R_\tau$ 
(the chain label $\alpha$ in $\R$ is dropped since the integration here 
is only over one chain), and
\begin{equation}
\Hp[\psi, \w]  = \frac{3}{2l^2}\sum_{\tau=1}^{N-1} \left(\R_{\tau+1}\!-\!\R_{\tau}\right)^2 
		+ \frac{i}{\Omega}\sum_{\kk\neq\kzero} \sum_{\tau=1}^{N} 
				 (\sigma_\tau \psi_\kk + \w_\kk)	e^{-i\kk\cdot\R_{\tau}} \; .
	\label{eq:Hp_of_Qp}
\end{equation}
The total interaction free energy involving the polymers
in the unit of Eq.~\ref{f_eq} is $-(l^3/\Omega)\ln\Zp$, which
we express as the sum of a density-fluctuation contribution
$\fpoly=-(l^3/\Omega)\ln\Zp^\prime$ and a mean-field contribution
$\fzero =-(l^3/\Omega)\ln\Zzero
= \frac{1}{2}\wtwo\rho_m^2$. The $\fzero$ term involves neither small 
ions nor electrostatic interactions because the excluded volumes of the
small ions are not considered beyond the incompressibility condition
in Eq.~\ref{eq_mix} and the solution system as a whole is neutral.

We evaluate $\Zp^\prime$ in Eq.~\ref{eq:Zp_H} perturbatively by 
expanding $\Eng[\psi, \w]$ to second order in density: 
\begin{equation}
\Eng[\psi, \w] \approx \frac{1}{2\Omega}\sum_{\kk\neq\kzero}
	\left\langle \psi_{-\kk} \; \w_{-\kk} \right|
\left(
\begin{array}{cc}
\vcol_k + \rho_m \Gss_\kk & \rho_m \Gds_\kk \\
\rho_m \Gds_\kk & {\wtwo}^{-1} + \rho \Gdd_\kk
\end{array}
\right)
\left|
\begin{array}{c}
\psi_\kk \\
\w_\kk
\end{array}
\right\rangle \; ,
	\label{eq:Hall_matrix}
\end{equation}
where $\Gdd_\kk$, $\Gss_\kk$, and $\Gds_\kk$ are monomer density-monomer 
density, charge-charge, and monomer density-charge correlation functions 
in $\kk$-space, $\langle \dots |$ and $|\dots\rangle$ are, respectively,
row and column vectors. $\Zp^\prime$ can then be calculated as a 
Gaussian integral to yield
\begin{equation}
\fpoly = -\frac{l^3 \ln\Zp^\prime}{\Omega} = \frac{l^3}{2}\int \frac{d^3 k}{(2\pi)^3}
	\ln\left[
		1 + \rho_m\left( \frac{\Gss_\kk}{\vcol_k} + \wtwo\Gdd_\kk \right)
		+ \frac{\wtwo}{\vcol_k} \rho_m^2 \left( \Gss_\kk\Gdd_\kk - \Gds_\kk^2 \right)
	\right] \; .
	\label{eq:fpoly}
\end{equation}
Evaluation of $\Gdd_\kk$, $\Gss_\kk$, and $\Gds_\kk$ requires knowledge
of the single-polymer $\Qp$ (Eq.~\ref{eq:Hall}), which in general
depends on the sequence charge pattern. fG-RPA makes the simplifying assumption
that $\Qp$ is that of Gaussian chains with a fixed $l$, i.e., assumes
that the second term in Eq.~\ref{eq:Hp_of_Qp} vanishes. As introduced
above, here we use a renormalized Kuhn length $l_1 = xl$ to better account
for the effects of interactions on $\Qp$ by making the improved approximation
\begin{equation}
\Qp \approx  \int \DD[\R] e^{-\Hp^0} \:\:;\:\:
\Hp^0 = \frac{3}{2l^2 x}\sum_{\tau=1}^{N-1}\left( \R_{\tau+1}-\R_\tau\right)^2 
\; .
	\label{eq:Hp0}
\end{equation}
Accordingly, the correlation functions 
in Eq.~\ref{eq:fpoly} are computed using $l_1$ instead of $l$:
\begin{equation}
\Gdd_\kk \to \Gdd_k^x =  \frac {1}{N} 
\bra{1} \hat{G}^x_k \ket{1} \; , \quad
\Gss_\kk \to \Gss_k^x =  \frac {1}{N} 
\bra{\sigma} \hat{G}^x_k \ket{\sigma} \; , \quad
\Gds_\kk \to \Gds_k^x =  \frac {1}{N}
\bra{\sigma} \hat{G}^x_k \ket{1} \; ,
	\label{eq:G_x}%
\end{equation}
where $\hat{G}^x_k$ is the $N\times N$ correlation matrix 
of the renormalized Gaussian chain 
with $[\hat{G}^x_k]_{\tau\mu} = \exp[ -(kl)^2x|\tau-\mu|/6 ]$, 
$\bra{1}$ and $\ket{1}$ are $N$-dimensional vectors 
with all elements equal to $1$. 

As emphasized above, the single $x$ variable here for end-to-end distance 
serves to provide an approximate account of
sequence specific effects in single-chain conformations.
A more accurate formalism that may be pursued in the future is to
consider $x$ as a function of specific residue pairs, i.e. $x$ $\rightarrow$
$x(\tau,\mu)$, so as to provide a structure factor that applies to all 
length scales as in the approach of Shen and Wang\cite{Shen2017}.

A variational approach similar to that in Sawle and Ghosh~\cite{Sawle2015} is 
applied to obtain a sequence-specific $x$ by first expressing
$\Hp$ in Eq.~\ref{eq:Hp_of_Qp} as $\Hp=\Hp^0+\Hp^1$ where 
$\Hp^0$ is given by Eq.~\ref{eq:Hp0} and $\Hp^1$ is the discrepany
in using the renormalized $\Hp^0$ to approximate $\Hp$. In general, a
partially optimized solution for $x$ may be obtained by minimizing the 
differences in averaged physical quantities computed using $\Hp$ versus 
those computed using $\Hp^0$, i.e., minimizing contributions from $\Hp^1$.
To simplify this calculation, we use, as in Ref.~\citen{Sawle2015}, the polymer
squared end-to-end distance $|\R_N-\R_1|^2$ as the physical quantity 
for the partial optimization of $x$.
The derivation proceeds largely as before~\cite{Sawle2015}, except the 
monomer-monomer interaction potential in Ref.~\citen{Sawle2015} is now
replaced by the effective field-field correlation function~\cite{Muthu1996}
\begin{equation}
U_{\rm eff}(\kk) \equiv  \sum_{\tau,\mu=1}^N
	\Big[  \sigma_\tau\sigma_\mu \avg{\psi_{-\kk}\psi_{\kk}} 
		+ \avg{\w_{-\kk}\w_{\kk}} 
		+ (\sigma_\tau \!+\! \sigma_\mu) \avg{\psi_{-\kk}\w_{\kk}} 
	 \Big] \; ,
\end{equation}
where $\langle \dots\rangle$ represents averaging over field 
configurations.
This analysis, the details of which are given in the appendix, 
leads to an equation that allows us to determine $x$:
\begin{equation}
1-\frac{1}{x} - \frac{N l^2}{18(N-1)} \int \!\!\frac{d^3 k}{(2\pi)^3} \frac{k^2  \Xee_k^x}{\det \Delta_k^x} = 0,
	\label{eq:x_sol_final}
\end{equation}
where $\Delta_k^x$ is the $2\times2$ matrix in Eq.~\ref{eq:Hall_matrix} with 
$\Gdd_\kk$, $\Gss_\kk$, and $\Gds_\kk$ replaced by their
renormalized $\Gdd_k^x$, $\Gss_k^x$, and $\Gds_k^x$ 
in Eq.~\ref{eq:G_x}. In the numerator of the integrand in
Eq.~\ref{eq:x_sol_final},
\begin{equation}
\Xee_k^x = \frac{\LGss_k^x}{\wtwo} + \vcol_k \LGdd_k^x + 
	\rho\left( \LGss_k^x \Gdd_k^x + \Gss_k^x\LGdd_k^x - 2\Gds_k^x \LGds_k^x \right),
\end{equation}
where 
\begin{equation}
\LGss_k^x =  \frac{1}{N}\bra{\sigma} \hat{L}_2\hat{G}^x_k\ket{\sigma} \; ,
\quad
\LGdd_k^x =  \frac{1}{N}\bra{1} \hat{L}_2\hat{G}^x_k\ket{1} \; ,
\quad
\LGds_k^x =  \frac{1}{N}\bra{\sigma} \hat{L}_2\hat{G}^x_k\ket{1} \; ,
\end{equation}
with $\hat{L}_2$ being an $N\times N$ matrix with 
$[\hat{L}_2]_{\tau\mu} = |\tau-\mu|^2$. Now, for any chosen
excluded-volume parameter $\wtwo$, 
$x$ can be solved as the only unknown in Eq.~\ref{eq:x_sol_final}. 
With $x$ determined, $\fpoly$ can be computed via Eq.~\ref{eq:fpoly} and
combined with the above expressions for $s$, $\fion$ and 
$\fzero$ to complete the free energy function in Eq.~\ref{f_eq} 
for our rG-RPA theory. Here we use $\wtwo=4\pi l^3/3$, which is 
about the $\sim l^3$ size of a monomer, in the applications below. 

We note that while $\wtwo>0$ (which disfavors collapsed
conformations) is required in the present formulation 
to solve for an effective Kuhn length,
the general trend predicted by our theory is not affected 
by reasonable variation around the $\wtwo=4\pi l^3/3$ value.

\section{Results}

\subsection{Salt-free rG-RPA unifies established LLPS trends of
both uniformly charged polyelectrolytes and neutral polyampholytes}

\begin{figure}
   \includegraphics[width=\columnwidth]{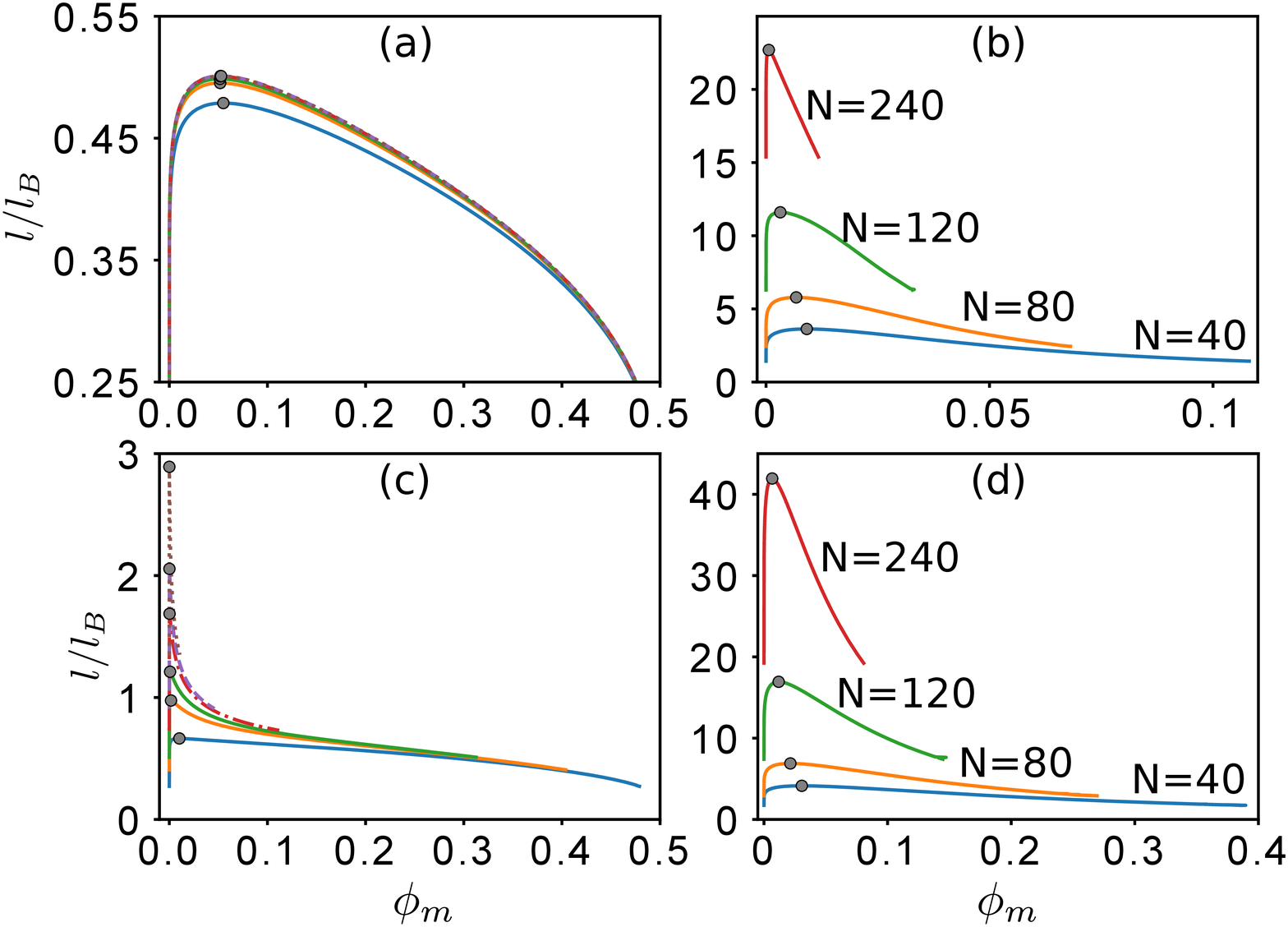} 
   \caption{Salt-free LLPS of polyelectrolytes 
and polyampholytes. rG-RPA (a and b, top panels) and 
fG-RPA (c and d, bottom panels) phase diagrams for $N=$ 10, 25, 40, 80, 
120, and 240 polyelectrolytes with charge sequences
$\sigma_\tau=-1$ for $\tau=1,2,\dots,N$ (a and c, left panels) 
and $N=40$, $80$, $120$, and $240$ 4-block polyampholytes with
charge sequences $\sigma_\tau=+1$ for $\tau=1,2,\dots,N/4$ and
$\tau=N/2+1,N/2+2,\dots,3N/4$, and  
$\sigma_\tau=-1$
for $\tau=N/4+1,N/4+2,\dots,N/2$ and $\tau=3N/4+1,3N/4+2,\dots,N$
(b and d, right panels).
Grey circles are critical points.
For the coexistence curves in (a and c), $N$ decreases from top to bottom,
with the $N=$ 80, 120, and 240 curves in (a) being nearly identical.}
   	\label{fig:PE-PA}
\end{figure}

We first illustrate the more general applicability of rG-RPA by comparing
rG-RPA and fG-RPA predictions for salt-free solutions of uniformly charged 
polyelectrolytes (fully charged homopolymers) and 4-block overall neutral 
polyampholytes of several different chain lengths (Fig.~\ref{fig:PE-PA}). 
As stated above, fG-RPA corresponds to setting $x=l_1/l=1$ and $\wtwo=0$
in rG-RPA. While fG-RPA is not identical to our earlier 
RPA~\cite{Lin2016} because fG-RPA subsumes the effects of small ions in
a screening potential for the polymers whereas our earlier RPA theory
treats the small ions and polymers on the same footing, 
both theories share the Gaussian-chain approximation
and their predicted trends are very similar, as will be illustrated
by examples below.


The rG-RPA-predicted 
critical point $((\phi_m)_{\rm cr}, 1/(\lb)_{\rm cr})$ 
in Fig.~\ref{fig:PE-PA}(a) for
polyelectrolytes is insensitive 
to chain length 
($(\lb)_{\rm cr}$ is critical Bjerrum length; 
$1/(\lb)_{\rm cr}$ is proportional to 
the critical temperature $T_{\rm cr}$). As $N$ increases, 
$\lim_{N\to\infty}1/(\lb)_{\rm cr}\approx 0.5$ and
$\lim_{N\to\infty}(\phi_m)_{\rm cr}\approx 0.05$.
These predictions are consistent with lattice-chain 
simulations~\cite{Orkoulas2003} and other 
theories~\cite{Jiang2001,Muthu2002,Budkov2015,Shen2017}. 
The fG-RPA predictions are drastically different, viz.,
$\lim_{N\to\infty}1/(\lb)_{\rm cr}\rightarrow\infty$ and 
$\lim_{N\to\infty}(\phi_m)_{\rm cr}\rightarrow 0$ (Fig.~\ref{fig:PE-PA}(c)). 
Thus, fG-RPA is limited as earlier RPA 
theories~\cite{Mahdi2000,Ermoshkin2003} and its predictions for 
polyelectrolytes are inconsistent with the aforementioned established 
results~\cite{Jiang2001,Muthu2002,Orkoulas2003,Budkov2015,Shen2017}. 
This comparison between rG-RPA and fG-RPA underscores the importance
of appropriately accounting for conformational heterogeneity in understanding 
polyelectrolyte LLPS and the effectiveness of using 
renormalized Kuhn lengths for the purpose.

Both rG-RPA and fG-RPA predict $1/(\lb)_{\rm cr}\rightarrow\infty$ and 
$(\phi_m)_{\rm cr}\rightarrow 0$ as $N\rightarrow\infty$ 
for the polyampholytes
(Fig.~\ref{fig:PE-PA}(b and d)). These results are 
consistent with simple RPA 
theory~\cite{Lin2016, Lin2017a}, a charged hard-sphere chain 
model~\cite{Jiang2006}, and lattice-chain simulations~\cite{Cheong2005}.
Not surprisingly, both rG-RPA and fG-RPA posit that the $T_{\rm cr}$'s
of polyelectrolytes are much lower than those of neutral polyampholytes
because direct electrostatic attractions exist for polyampholytes 
but effective attractions among polyelectrolytes can only be mediated by
counterions.

For the polyampholytes, rG-RPA (Fig.~\ref{fig:PE-PA}(b)) predicts lower 
$T_{\rm cr}$'s than fG-RPA (Fig.~\ref{fig:PE-PA}(d)).
With a more accurate treatment of single-chain conformational dimensions,
rG-RPA should entail more compact isolated single-chain conformations for block 
polyampholytes, resulting in less accessibility of the charges for
interchain cohesive interactions and therefore a weaker---but physically
more accurate---LLPS propensity.

Notably, the fG-RPA-predicted phase boundaries of both polyelectrolytes 
and polyampholytes exhibit an inverse S-shape phase boundaries 
(the condensed-phase part of the coexistence curves concave upward;
Fig.~\ref{fig:PE-PA}(c and d)). In contrast, rG-RPA predicts that only
polyampholytes have inverse S-shape phase boundaries (Fig.~\ref{fig:PE-PA}(b)), 
whereas polyelectrolytes phase boundaries convex upward with a
relatively flat $\phi_m$ dependence around the critical points 
(Fig.~\ref{fig:PE-PA}(a)). This conspicuous difference between the 
rG-RPA-predicted phase boundaries of polyampholytes and polyelectrolytes 
is consistent with explicit-chain simulations~\cite{Orkoulas2003,Das2018b}.

\subsection{Salt-free rG-RPA account of pH-dependent LLPS}

\begin{figure}
   \includegraphics[width=\columnwidth]{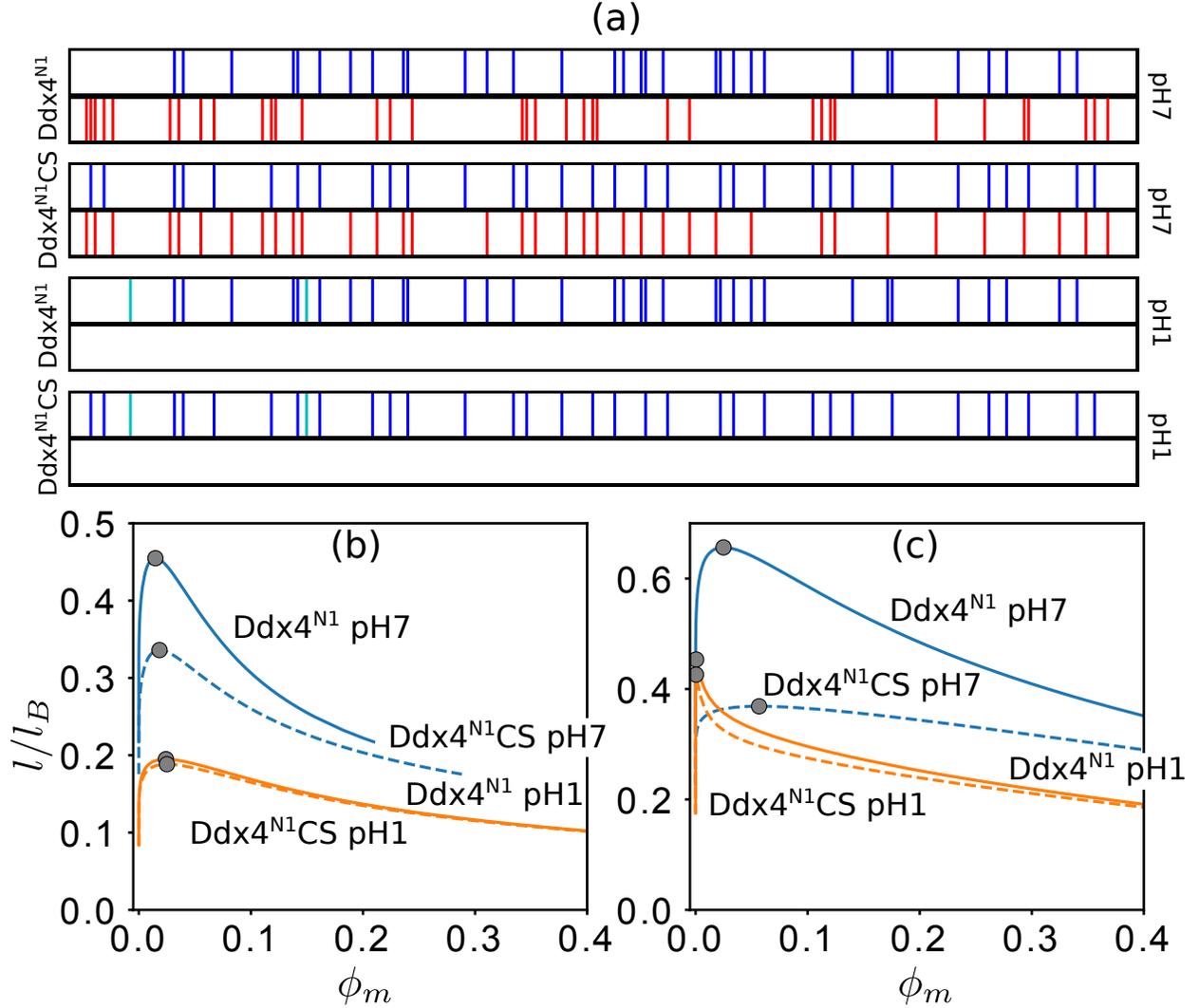} 
   \caption{LLPS at neutral and acidic pH. (a) Charge 
sequences of $\DdxN{}$ and $\DdxNCS{}$ (blue/cyan: 
$+1$, red: $-1$, white: $0$) and
their (b) rG-RPA and (c) fG-RPA phase~diagrams.}
   	\label{fig:Ddx4s}
\end{figure}

To address pH dependence under salt-free conditions, we apply rG-RPA to 
an example of a near-neutral polyampholyte under neutral pH, namely the 
N-terminal IDR of the DEAD-box helicase Ddx4 (IDR denoted as 
Ddx4$^{\rm N1}$) and its charge-scrambled variant Ddx4$^{\rm N1}$CS which 
has the same amino acid composition as Ddx4$^{\rm N1}$ by a different 
sequence charge pattern~\cite{Nott2015}. The sequences are studied at 
neutral and acidic pH. We refer to the resulting charge patterns as
(in obvious notation) $\DdxNn$, $\DdxNCSn$, $\DdxNl$, and $\DdxNCSl$, 
where pH7 and pH1 are approximate pH values symbolizing neutral and acidic
conditions. For the pH7 sequences, each of the 24 arginines (R) and 8 
lysines (K) of Ddx4$^{\rm N1}$ and Ddx4$^{\rm N1}$CS is assigned a $+1$ 
charge, each of the 18 aspartic acids (D) and 18 glutamic acids (E) is 
assigned a $-1$ charge, and the 2 histidines (H) carry zero charge.
For the pH1 sequences, because the pH is lower than the pKa of the acidic 
amino acids (3.71 for D and 4.15 for E), they are not ionized and thus
carry zero charge but each K or R or H ($p{\rm K_H} = 6.04$) carries 
a $+1$ charge (Fig.~\ref{fig:Ddx4s}(a), K, R in blue; H in cyan). 
Thus, $\DdxNn$ and $\DdxNCSn$ are near-neutral polyampholytes whereas 
$\DdxNl$ and $\DdxNCSl$ are polyelectrolytes, although these four 
sequences---unlike those in Fig.~\ref{fig:PE-PA}---contains also many 
uncharged monomers.


Fig.~\ref{fig:Ddx4s}(b) indicates that the rG-RPA-predicted $T_{\rm cr}$ 
is much lower under acidic than under neutral conditions, and that the 
$T_{\rm cr}$ of $\DdxN{}$ is always higher than that of $\DdxNCS{}$ under both
pH conditions, underscoring that sequence-specific effects influence the 
LLPS of not only neutral and nearly-neutral 
polyampholytes~\cite{Lin2016, Lin2017a, Lin2017b, Das2018a, Das2018b} 
but also polyelectrolytes. Intriguingly, inverse S-shaped coexistence curves
are seen in Fig.~\ref{fig:Ddx4s}(b) not only for neutral pH (blue curves) 
but also for acidic pH (orange curves). 
This feature is characteristic of polyampholytes 
(Fig.~\ref{fig:PE-PA}(b)) but not uniformly charged 
polyelectrolytes (Fig.~\ref{fig:PE-PA}(a)). This result suggests
that inverse S-shaped phase boundaries can arise in general from a 
heterogeneous sequence charge pattern because it leads to the simultaneous 
presence of both attractive and repulsive interchain interactions (which
can be counterion-mediated in the case of polyelectrolytes)
and therefore allows for condensed-phase configurations with lower 
densities~\cite{Das2018b}.

\begin{figure}
   \includegraphics[width=\columnwidth]{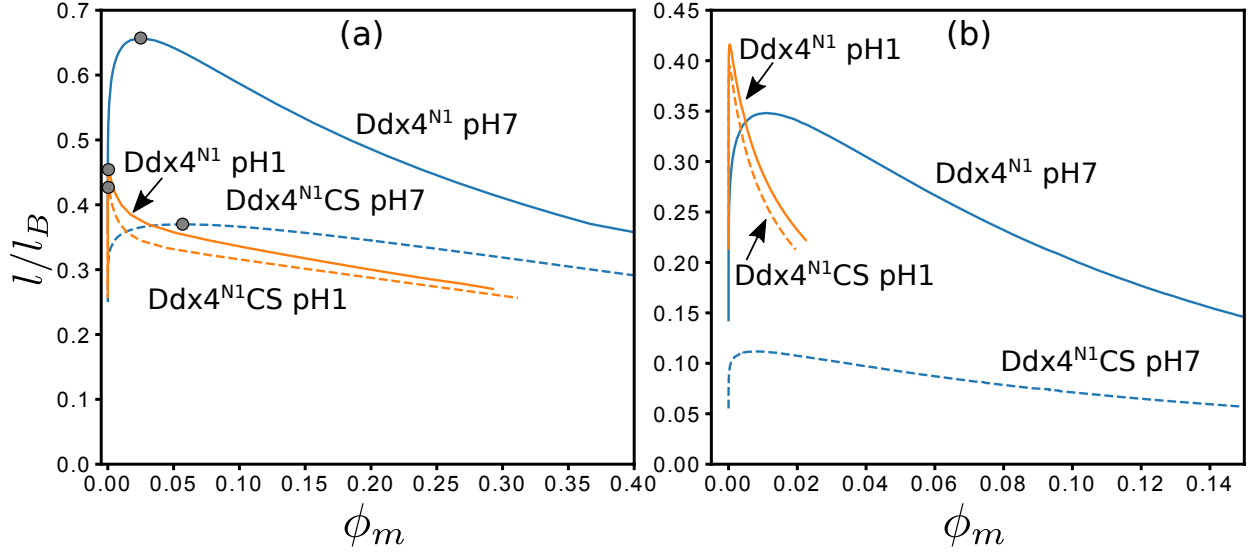} 
   \caption{Simple RPA~\cite{Lin2016,Lin2017a} salt-free phase diagrams 
for the four Ddx4 sequences in Fig.~\ref{fig:Ddx4s}(a). (a) Phase diagrams
computed using the Coulomb potential in Fourier space, $U_k = 4\pi\lb/k^2$, 
are very similar to the fG-RPA phase diagrams in Fig.~\ref{fig:Ddx4s}(c). 
(b) Phase diagrams computed using a Coulomb potential with a short-range 
cutoff, $U_k = 4\pi\lb/[k^2(1+(kl)^2)]$;
the same potential used in our previous 
simple-RPA studies~\cite{Lin2016, Lin2017a, Lin2017b, Lin2017c, Brady2017}. 
This Coulomb potential with a short-range cutoff predicts that the 
two pH1 sequences have critical temperatures even higher than that 
of wildtype Ddx4 at pH7. This prediction, however, contradicts the physical 
intuition that polyelectrolytes should have lower phase separation
propensities than neutral or near-neutral polyampholytes of the same
chain length. }
   	\label{fig:Ddx4s_simpleRPA_saltfree}
\end{figure}

As a control, fG-RPA results are shown in Fig.~\ref{fig:Ddx4s}(c).
In contrast to rG-RPA, fG-RPA predicts that the $l/(\lb)_{\rm cr}$ value
(proportional to $T_{\rm cr}$) of both Ddx4$^{\rm N1}$ and Ddx4$^{\rm N1}$CS 
at low pH is higher than that of Ddx4$^{\rm N1}$CS at neutral pH, and that
the critical volume fractions at low pH are significantly lower than those
at neutral pH. Although these differences between fG-RPA and rG-RPA 
predictions for the Ddx4 IDR remain to be conclusively tested by experiment, 
the low-pH fG-RPA phase diagrams here (orange curves in Fig.~\ref{fig:Ddx4s}(c))
share similar features with the fG-RPA phase diagrams for polyelectrolytes
in Fig.~\ref{fig:PE-PA}(c) which, as discussed above, are at odd with
trends observed in prior theories and experiments. 
The fG-RPA results and those obtained 
using our earlier, simple formulation of RPA~\cite{Lin2016} 
are very similar (Fig.~\ref{fig:Ddx4s_simpleRPA_saltfree}).


\subsection{Salt-free rG-RPA rationalizes pH-dependent LLPS of IP5} 
\begin{figure}[h!]
   \includegraphics[width=0.75\columnwidth]{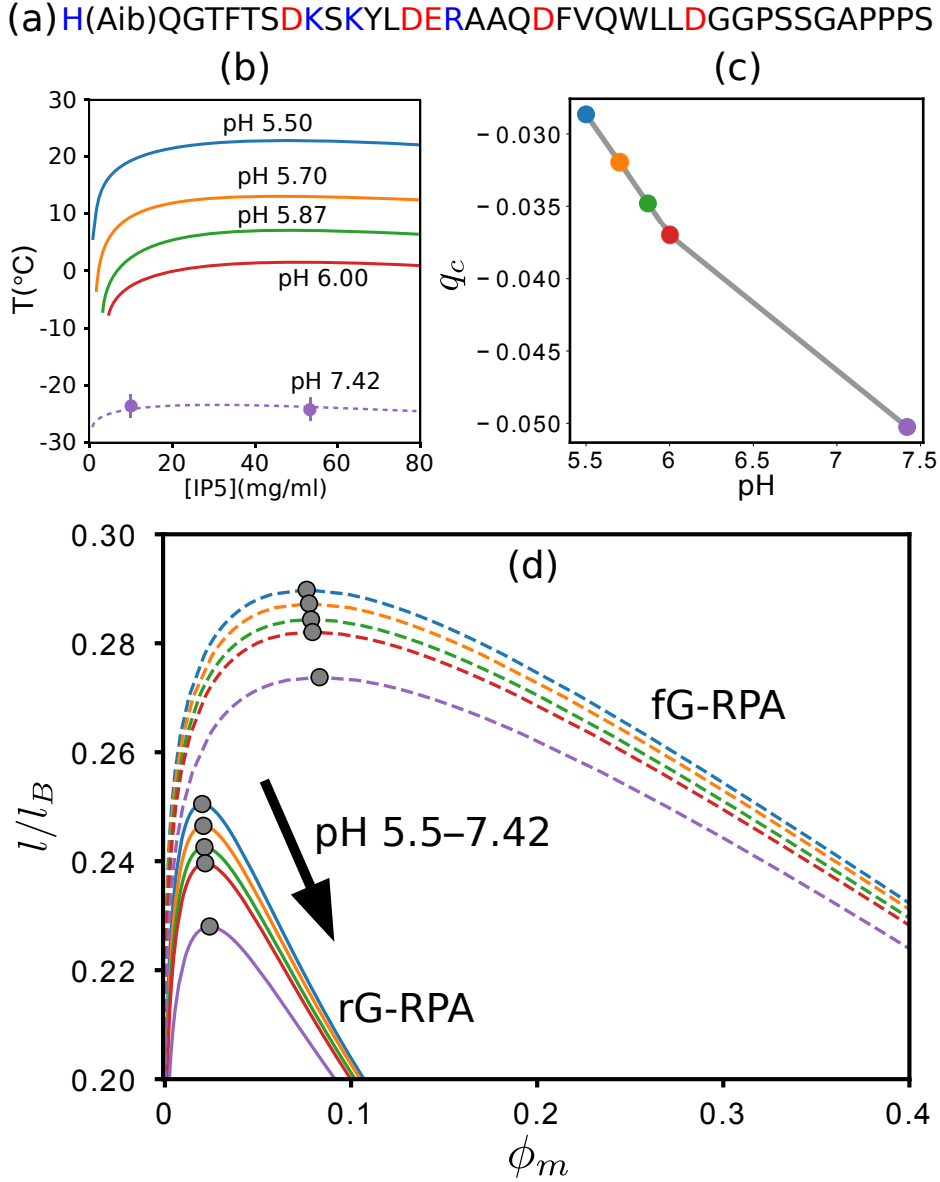} 
   \caption{LLPS of IP5. (a) The IP5 sequence, where
basic and acidic residues are in blue and red, respectively; (Aib) is 
the non-proteinogenic amino acid $\alpha$-methylalanine~\cite{Wang2017}. 
(b) Experimental pH-dependent phase diagrams of IP5 based on the data in 
Fig.~4 of Ref.~\citen{Wang2017}; anti-freeze was used to obtain some
of the low-$T$ results~\cite{Wang2017}.
(c) Net charge per residue, $\pc$, of IP5.
(d) Phase diagrams predicted by rG-RPA (solid curves) and fg-RPA 
(dashed curves).}
   	\label{fig:IP5}
\end{figure}

We now utilize our theory to rationalize part of the 
experimental pH-dependent LLPS trend of the lyophilized 39-residue peptide 
IP5~\cite{Wang2017},
the isoelectric point of which is pH = 4.4
(Fig.~\ref{fig:IP5}(a and b))~\cite{CRCbook}. 
The pH-dependent charge $\sigma$ of a basic or acidic residue
is computed~\cite{Ghosh2009} here by 
\begin{equation}
\sigma = \pm \frac{10^{\pm({\rm pK}_{\rm a}-{\rm pH})}}{1+10^{\pm({\rm pK}_{\rm a}-{\rm pH})}} \; ,
	\label{eq:ch_pH-dept}
\end{equation}
where the $+$ and $-$ signs in the $\pm$ signs above apply
to the basic (R, K, H)
and acidic (D, E) residues, respectively. Standard ${\rm pK}_{\rm a}$ 
values~\cite{CRCbook}, viz., R: $12.10$, K: $10.67$, H: $6.04$,
D: $3.71$, and E: $4.15$, are used in Eq.~\ref{eq:ch_pH-dept}
to construct pH-dependent charge sequences of IP5 (Fig.~\ref{fig:IP5}(c)).

The rG-RPA- and fG-RPA-predicted IP5 phase boundaries for the experimental
studied pH values are shown in Fig.~\ref{fig:IP5}(d). Both theories
predict a lower $l/(\lb)_{\rm cr}\approx 0.2$--$0.3$ than the experiment 
$l/(\lb)_{\rm cr}\approx 0.5$. Physically, this is not surprising, 
as has been addressed in previous RPA studies~\cite{Lin2016}, 
because non-electrostatic cohesive interactions are neglected here.
Nonetheless, consistent with experiment, both theories
posit that LLPS propensity decreases with increasing pH. Moreover,
the rG-RPA-predicted critical volume fraction $(\phi_m)_{\rm cr}\approx 
0.020$--$0.024$
is reasonable in view of the experimental value 
of $\approx 0.036$ (Ref.~\citen{Wang2017}), indicating once again that
rG-RPA is superior to fG-RPA as the latter predicts much higher 
$(\phi_m)_{\rm cr}$'s.

\subsection{Salt-dependent rG-RPA for heteropolymeric charge sequences}

\begin{figure}[h!]
   \includegraphics[width=\columnwidth]{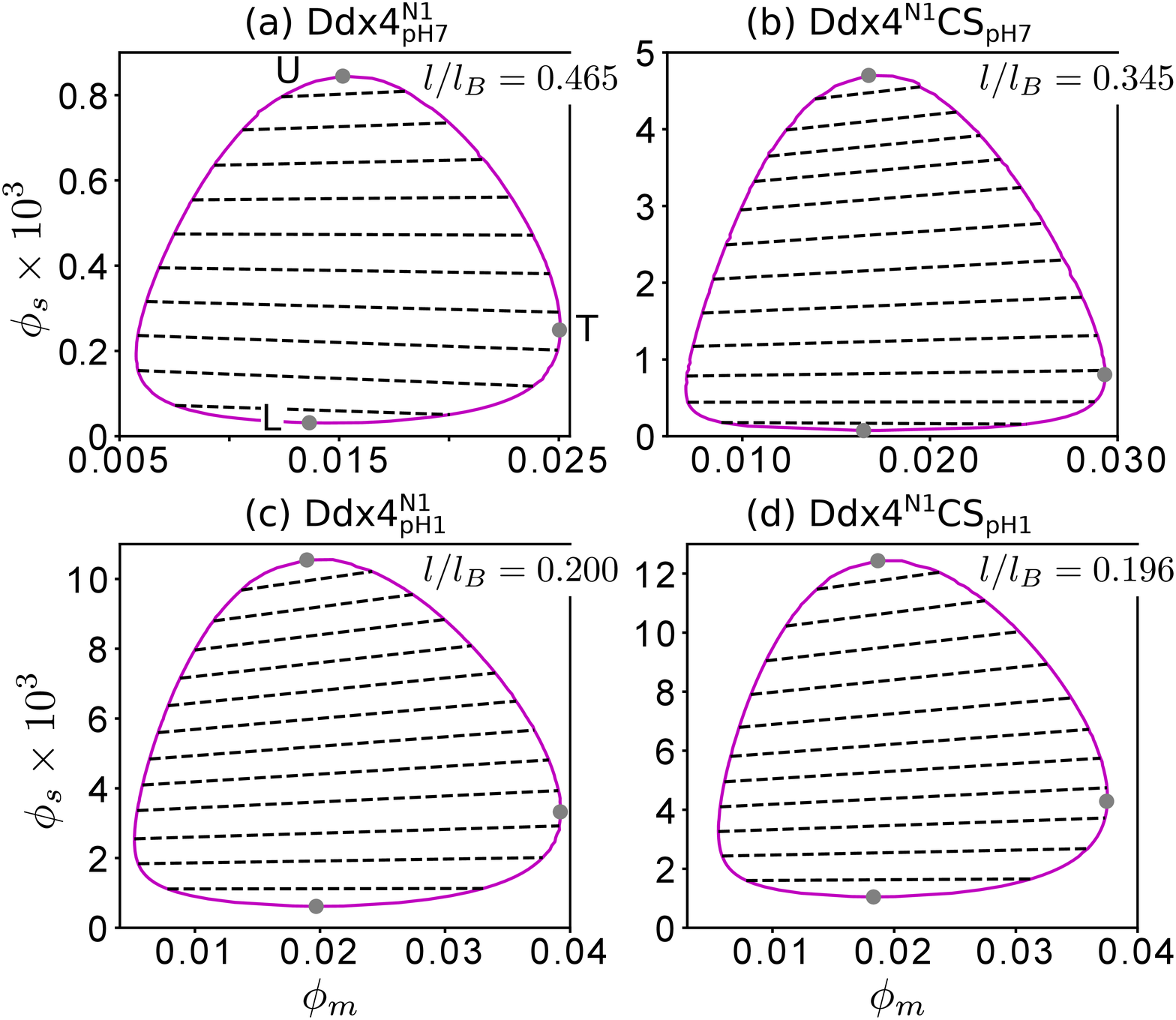} 
   \caption{IDR-salt binary phase diagrams of two Ddx4 variants at low and high
pH. Results are for $l/\lb \gtrsim l/(\lb)_{\rm cr}$, where the salt-free 
$1/(\lb)_{\rm cr}$ equals $0.455$ for $\DdxNn$ (a), $0.336$ for $\DdxNCSn$ (b),
$0.195$ for $\DdxNl$ (c), and $0.188$ for $\DdxNCSl$ (d). The $\phi_s$ values 
of the grey circles in (a)--(d) are $(\phi_s)_{\rm cr}^{\rm U}$, 
$(\phi_s)^{\rm T}$, or $(\phi_s)_{\rm cr}^{\rm L}$, as indicated by 
U, T, and L in (a).
   }
   \label{fig:Ddx4_all_salt_lb_high}
\end{figure}

In view of the superiority 
of rG-RPA over fG-RPA, only rG-RPA is used below. 
We consider the four charge sequences
in Fig~\ref{fig:Ddx4s}(a) as examples and restrict 
attention to monovalent salt and counterions ($\zs=\zc=1$). In 
experiments we conducted for this study
using described methods~\cite{Brady2017}, 
no Ddx4$^{\rm N1}$ LLPS was observed in salt-free solution at
room temperature; yet Ddx4$^{\rm N1}$ at room temperature is 
known~\cite{Nott2015, Brady2017}
to phase separate with 100 mM NaCl and that
LLPS propensity decreases when [NaCl] is increased to 300 mM.
These findings suggest that, similar
to LLPS of uniformly charged 
polyelectrolytes~\cite{Eisenberg1959,Sabbagh2000,Vivek},
salt dependence of heteropolymer LLPS is non-monotonic 
at temperatures slightly higher than the salt-free $T_{\rm cr}$
and therefore such temperatures are of particular interest. For this 
reason, we apply rG-RPA to compute IDR-salt binary phase diagrams
of $\DdxNn$, $\DdxNCSn$, $\DdxNl$, and $\DdxNCSl$ 
(Fig.~\ref{fig:Ddx4_all_salt_lb_high}), each at an $l/\lb$ value slightly 
higher than the sequence's salt-free $l/(\lb)_{\rm cr}$
in Fig~\ref{fig:Ddx4s}(b). 

As expected, all binary phase diagrams in Fig.~\ref{fig:Ddx4_all_salt_lb_high}
exhibit non-monotonic salt dependence.
In general, at temperatures above the salt free critical temperature, i.e. 
$l/\lb \gtrsim$ salt-free $l/(\lb)_{\rm cr}$, when sufficient salt is 
added to the salt-free homogeneous solution, LLPS is triggered 
at $\phi_s = (\phi_s)^{\rm L}_{\rm cr}$. 
Adding more salt beyond $(\phi_s)^{\rm L}_{\rm cr}$ enhances LLPS
in that a wider range of overall $\phi_m$ falls within the LLPS regime, 
until a turning point $(\phi_s)^{\rm T}$ is reached. 
Beyond that, adding more salt (increasing $\phi_s$ above $(\phi_s)^{\rm T}$)
reduces LLPS (the phase-separated range of $\phi_m$ narrows). 
LLPS is impossible for the given temperature
when salt concentration is increased above
an upper critical point $(\phi_s)^{\rm U}_{\rm cr}$. 

Despite these qualitative commonalities, there are significant sequence-dependent differences. Notably,
at neutral pH, the range of salt concentrations that can induce LLPS is 
much narrower for $\DdxNn$ ($\phi_s \lesssim 0.00085$, 
Fig.~\ref{fig:Ddx4_all_salt_lb_high}(a)) than for 
$\DdxNCSn$ ($\phi_s \lesssim 0.005$, Fig.~\ref{fig:Ddx4_all_salt_lb_high}(b)). 
However, the ranges of LLPS-inducing salt concentrations at low pH
for $\DdxNl$ and $\DdxNCSl$ are similar 
($\phi_s \lesssim 0.01$, Fig.~\ref{fig:Ddx4_all_salt_lb_high}(c and d)),
and their $(\phi_s)^{\rm L}_{\rm cr}$ and $(\phi_s)^{\rm U}_{\rm cr}$ 
are significantly larger than those at neutral pH.

\begin{figure}
   \includegraphics[width=\columnwidth]{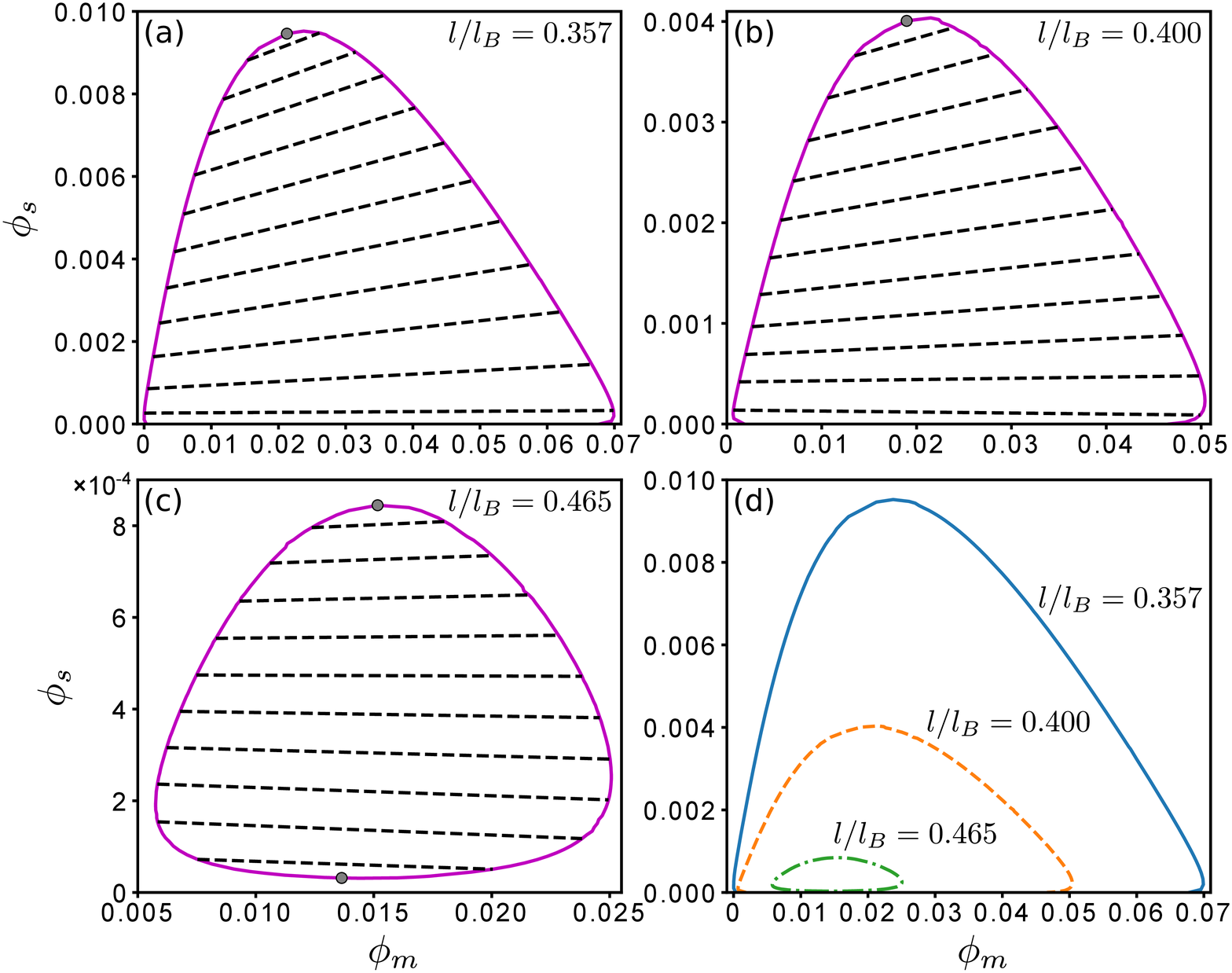} 
   \caption{Polymer-salt coexistence phase diagrams of $\DdxNn$ 
at the $l/\lb$ values indicated. The salt-free critical value of $l/\lb$
is $l/(\lb)_{\rm cr}=0.455$.
    Top grey circles in (a), (b), and (c) provide the upper critical salt 
concentrations $(\phi_s)^{\rm U}_{\rm cr}$, whereas the bottom grey circle
in (c) provides the lower critical concentration $(\phi_s)^{\rm L}_{\rm cr}$
(see discussion in main text).
    Each dashed line in (a)--(c) is a tie line connecting a pair of 
coexistent phases. The three phase boundaries in (a)--(c) are compared 
in (d). }
   	\label{fig:Ddx4_N1_pH7_salt}
\end{figure}

\begin{figure}
   \includegraphics[width=\columnwidth]{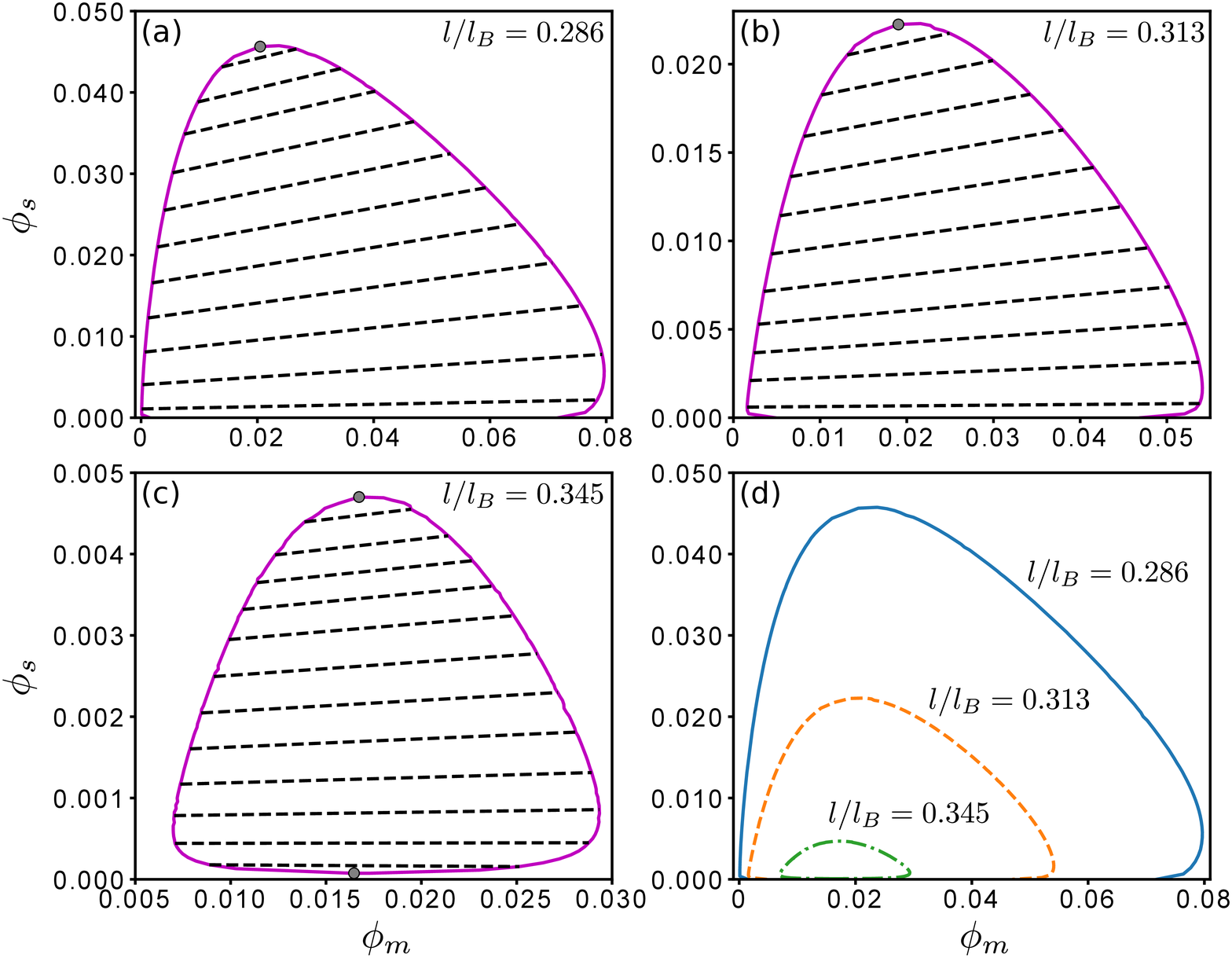} 
\caption{Polymer-salt coexistence phase diagrams of $\DdxNCSn$
at the $l/\lb$ values indicated. The salt-free critical value of $l/\lb$
is $l/(\lb)_{\rm cr}=0.336$.
    Top grey circles in (a), (b), and (c) provide the upper critical salt
concentrations $(\phi_s)^{\rm U}_{\rm cr}$, whereas the bottom grey circle
in (c) provides the lower critical concentration $(\phi_s)^{\rm L}_{\rm cr}$.
    Each dashed line in (a)--(c) is a tie line connecting a pair of 
coexistent phases. The three phase boundaries in (a)--(c) are compared 
in (d). }
   	\label{fig:Ddx4_N1_CS_pH7_salt}
\end{figure}

\begin{figure}
   \includegraphics[width=\columnwidth]{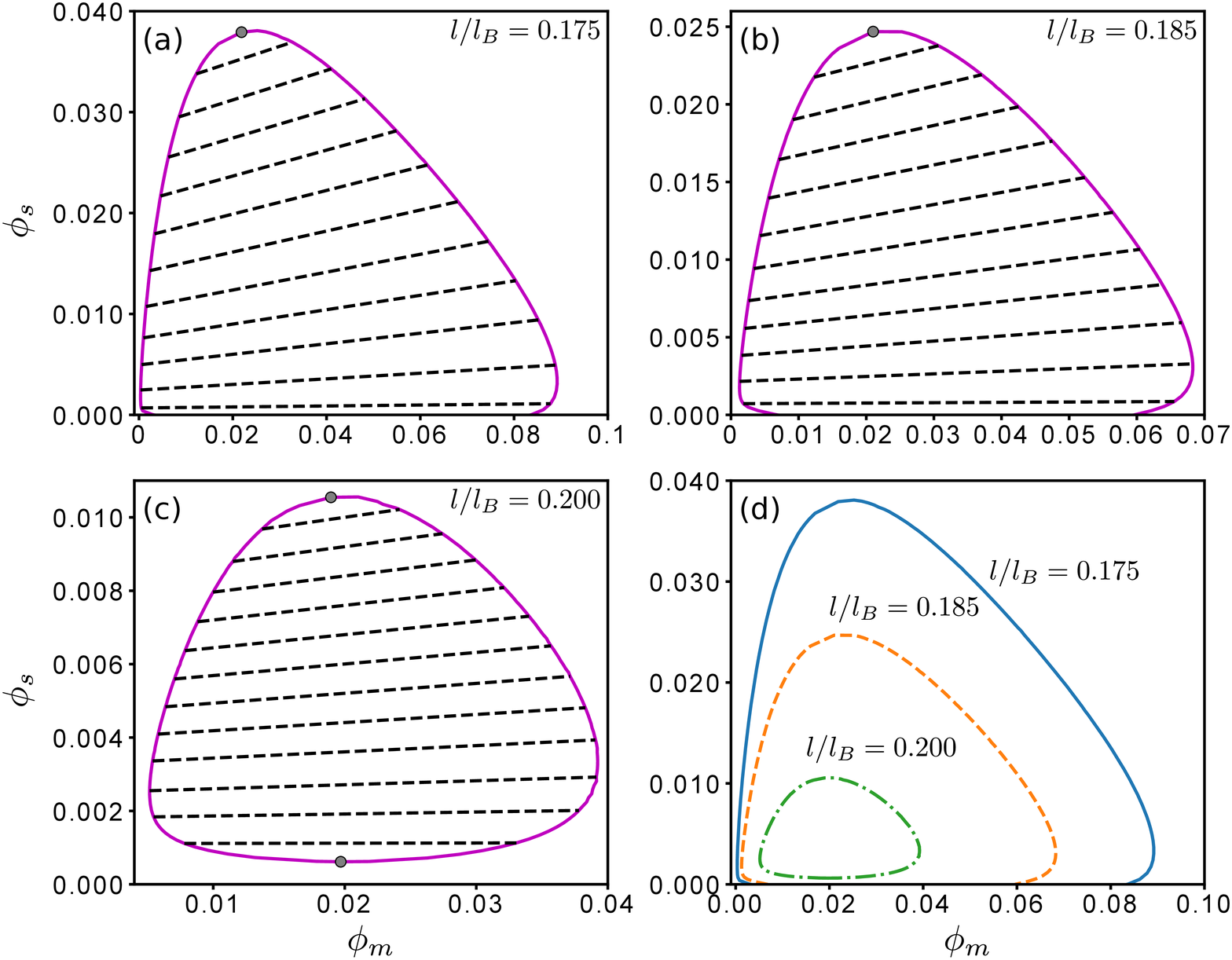} 
\caption{Polymer-salt coexistence phase diagrams of $\DdxNl$
at the $l/\lb$ values indicated. The salt-free critical value of $l/\lb$
is $l/(\lb)_{\rm cr}=0.195$.
    Top grey circles in (a), (b), and (c) provide the upper critical salt
concentrations $(\phi_s)^{\rm U}_{\rm cr}$, whereas the bottom grey circle
in (c) provides the lower critical concentration $(\phi_s)^{\rm L}_{\rm cr}$.
    Each dashed line in (a)--(c) is a tie line connecting a pair of
coexistent phases. The three phase boundaries in (a)--(c) are compared
in (d). }
   	\label{fig:Ddx4_N1_pH1_salt}
\end{figure}

\begin{figure}
   \includegraphics[width=\columnwidth]{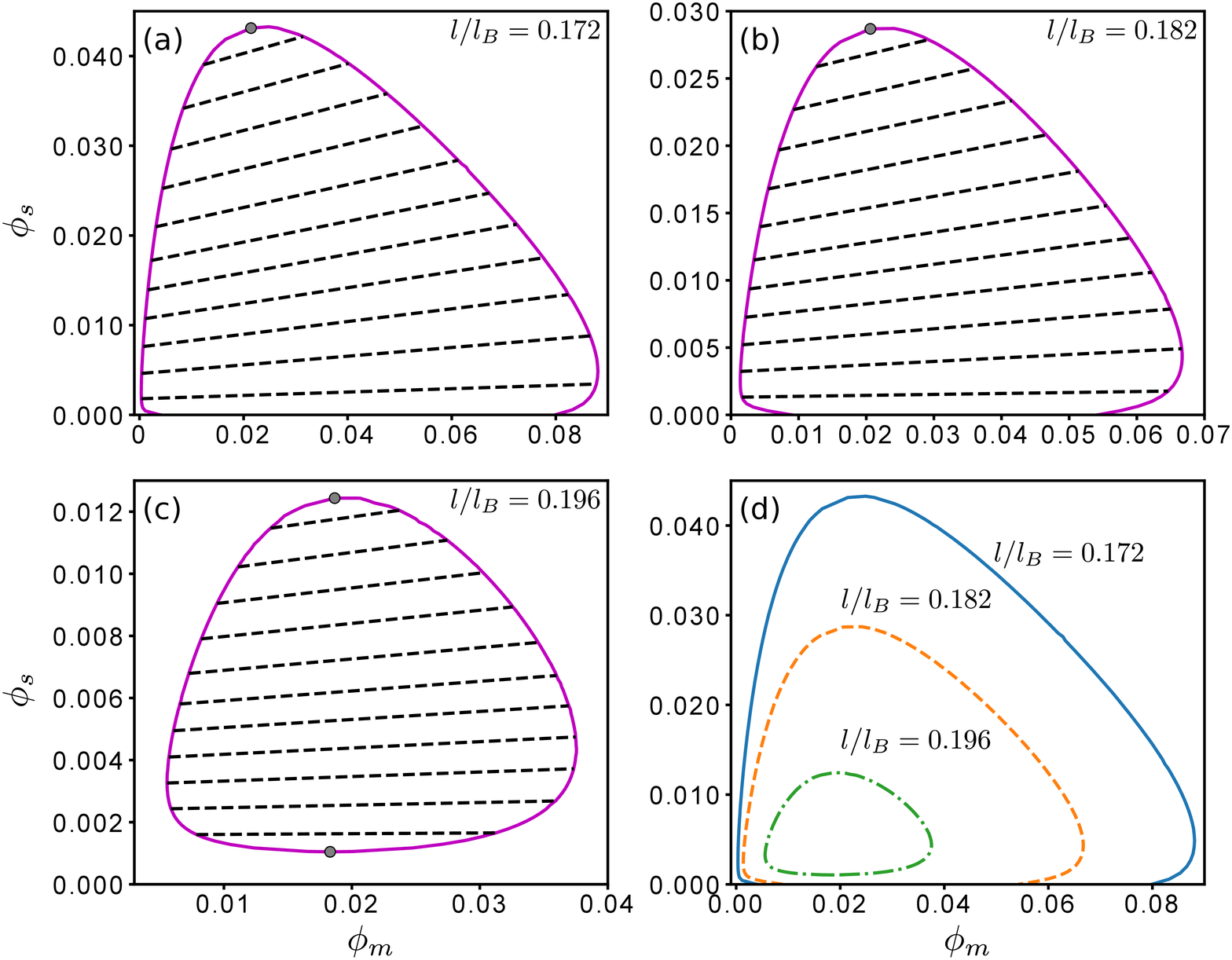} 
\caption{Polymer-salt coexistence phase diagrams of $\DdxNCSl$
at the $l/\lb$ values indicated. The salt-free critical value of $l/\lb$
is $l/(\lb)_{\rm cr}=0.188$.
    Top grey circles in (a), (b), and (c) provide the upper critical salt
concentrations $(\phi_s)^{\rm U}_{\rm cr}$, whereas the bottom grey circle
in (c) provides the lower critical concentration $(\phi_s)^{\rm L}_{\rm cr}$.
    Each dashed line in (a)--(c) is a tie line connecting a pair of
coexistent phases. The three phase boundaries in (a)--(c) are compared
in (d). }
   	\label{fig:Ddx4_N1_CS_pH1_salt}
\end{figure}

Next we explore these trends at temperatures below the salt-free $T_{\rm cr}$. 
Figs.~\ref{fig:Ddx4_N1_pH7_salt}--\ref{fig:Ddx4_N1_CS_pH1_salt} present salt-polymer
phase diagrams for four Ddx4 sequences (both wild type and charge scrambled sequences at neutral and acidic pH)
at three different temperatures. Panels (a) and (b) in these figures show phase diagrams at temperatures below the respective 
salt free $T_{\rm cr}$ for the given sequence, while panel (c) is at a temperature above salt free $T_{\rm cr}$.
The three phase diagrams are compared in panel (d) for a given sequence. These figures reveal trends for 
$l/\lb \gtrsim$ salt-free $l/(\lb)_{\rm cr}$ (above the salt free critical temperature) are largely in line with 
behaviors at temperatures below the salt-free $T_{\rm cr}$. The only difference is for $l/\lb <$ salt-free $l/(\lb)_{\rm cr}$, 
$(\phi_s)^{\rm L}_{\rm cr}=0$.  For $l/\lb <$ salt-free $l/(\lb)_{\rm cr}$, temperatures for different sequences 
were chosen such that the maximum $\phi_m$ range of LLPS are similar among the sequences (as in Fig.~\ref{fig:Ddx4_all_salt_lb_high}). 
With this choice of temperature constraint, when the IDR-salt phase diagrams for different sequences (Figs.~\ref{fig:Ddx4_N1_pH7_salt}--\ref{fig:Ddx4_N1_CS_pH1_salt}) 
are compared, we note that $(\phi_s)_{\rm cr}^{\rm U}$ and $(\phi_s)^{\rm T}$ of $\DdxNn$ are much smaller than those of $\DdxNCSn$. Furthermore,
$(\phi_s)_{\rm cr}^{\rm U}$, $(\phi_s)^{\rm T}$ of these two pH7 sequences are much smaller than those
of the two pH1 sequences. Thus, we conclude that $\DdxNn$ is more sensitive to salt than $\DdxNCSn$, and 
both are more salt-sensitive than $\DdxNl$ and $\DdxNCSl$.
Metrics other than $(\phi_s)^{\rm T}$ can also be used to determine salt sensitivity.
For example, the low-$\phi_m$ turning point (e.g., at $\phi_m\approx 0.006$, $\phi_s\approx 0.16$ in Fig.~\ref{fig:Ddx4_all_salt_lb_high}(a), unlabeled)
with a $\phi_s$ value similar to that of $(\phi_s)^{\rm T}$ may be used to characterize salt sensitivity. The resulting trend
is similar to the one gleaned from the turning point, $(\phi_s)^{\rm T}$.

The existence of a $(\phi_s)^{\rm L}_{\rm cr}>0$ in 
Fig.~\ref{fig:Ddx4_all_salt_lb_high}(a)
is consistent with our experimental observation that Ddx4$^{\rm N1}$
does not phase separate with [NaCl] $<$ 15--20 mM at pH 6.5, 
25$^\circ$C ($l/\lb=0.529$), and 5mM Tris. Other predictions of our theory 
remain to be tested. Of particular interest is the slopes of the
tie lines in Fig.~\ref{fig:Ddx4_all_salt_lb_high}(a) and (b) that 
change from negative to
positive as $\phi_s$ increases, indicating that salt ions and the
heteropolymeric IDRs partially
exclude each other in low-salt but partially coalesce 
in high-salt solutions at neutral pH. This intriguing feature was not 
encountered in solutions of either a single species of uniformly-charged
or two species of oppositely-charged 
homopolymers~\cite{Moreira2001, Zhang2016, Lytle2016, Lytle2017, SingMacro2017,Shen2018,TirrellMacro2018}.
In contrast, the tie-line slopes in 
Fig.~\ref{fig:Ddx4_all_salt_lb_high}(c) and (d) are all positive, indicating
that salt ions and the heteropolymeric IDRs always partially coalesce 
under acidic conditions.

\subsection{Salt-dependent rG-RPA is consistent with established
trends in LLPS of homopolymeric, uniformly charged polyelectrolytes}
\begin{figure}[h!]
   \includegraphics[width=\columnwidth]{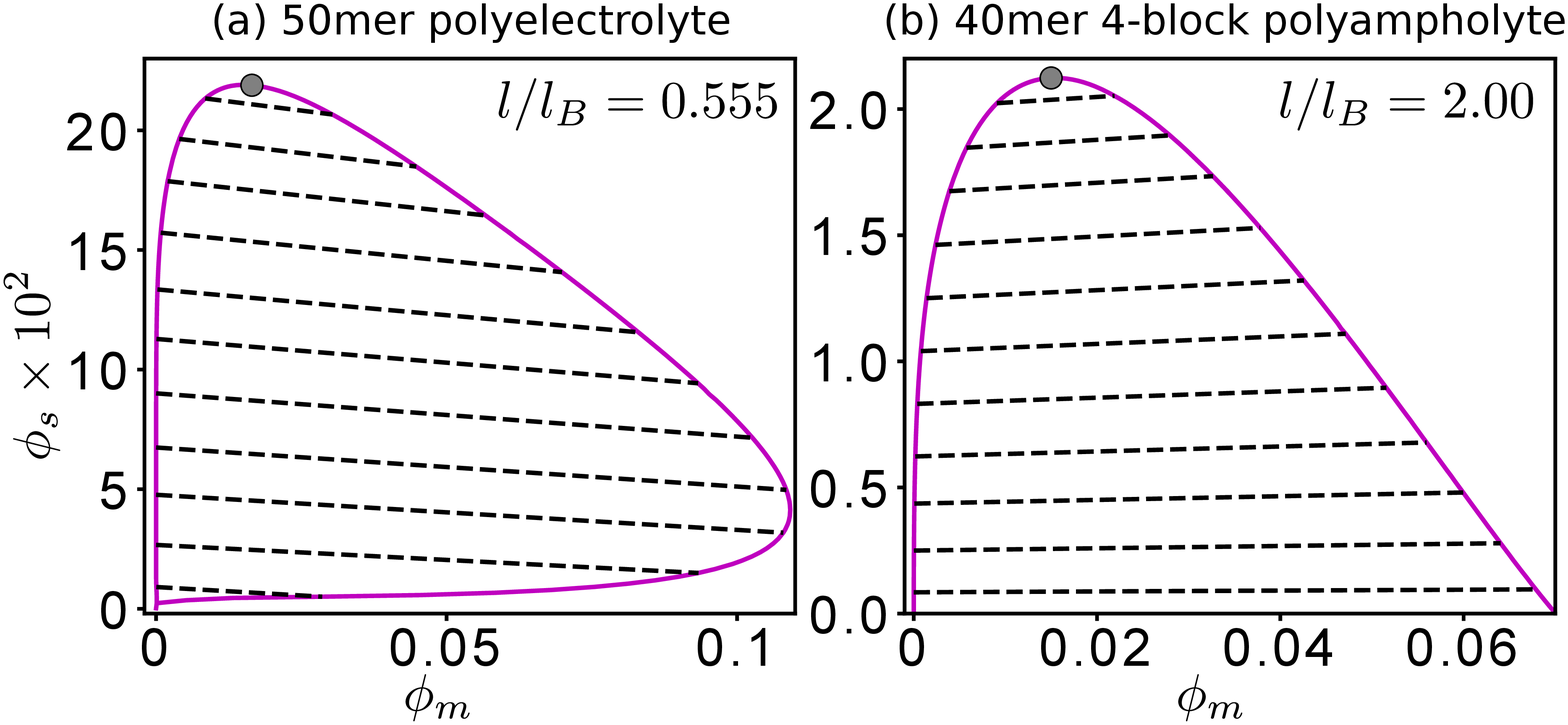} 
   \caption{Salt-dependent LLPS of 
polyelectrolytes and polyampholytes. rG-RPA phase diagrams 
for (a) an $N=50$ homopolymer with monomer charge $=-1$, and (b) 
the $N=40$ 4-block polyampholyte in Fig.~\ref{fig:PE-PA}. Note that
salt-free $l/(\lb)_{\rm cr}=$ $0.5$ for (a) and $=3.63$ for (b).
$(\phi_s)_{\rm cr}^{\rm U}$ is given by the grey circle.
An unmarked $\phi_s$ $=(\phi_s)_{\rm cr}^{\rm L}>0$ exists for (a)
but not for (b).
   }
   \label{fig:50E_KE4b}
\end{figure}

Our model predicts salt and polymers coalesce for $\DdxNl$ and $\DdxNCSl$ (Fig.~\ref{fig:Ddx4_all_salt_lb_high} c and d). These sequences are examples of 
non-uniformly charged polyelectrolytes. However, these results are in contrary to experiment and theory on uniformly charged polyelectrolytes that suggest salt ions and 
polymers tend to exclude each other, leading to tie lines with negative slopes in the polymer-salt phase diagrams~\cite{Moreira2001,Zhang2016, Shen2018,Eisenberg1959}. 
We test the ability of our model to reproduce this established trend by computing salt-polymer phase diagrams for uniformly charged polymers (Fig.~\ref{fig:50E_KE4b}(a)). 
The established feature is captured by our new theory, as the slopes of all tie lines are negative in Fig.~\ref{fig:50E_KE4b}(a). Furthermore, consistent with literature reports on 
uniformly charged homopolymers (homopolyelectrolytes)~\cite{Eisenberg1959,Sabbagh2000,Vivek}, with addition of salt, rG-RPA predicts a one-to-two phase transition in the low salt regime 
as well as a two-to-one phase transition in the high salt regime.
For comparison, Fig.~\ref{fig:50E_KE4b}(b) is the 
phase diagram of an overall neutral polyampholytes at a temperature 
substantially lower than the salt-free $T_{\rm cr}$ with all tie lines 
having positive slopes. A recent field theory simulation study of an
overall neutral diblock polyampholyte also found tie lines with slightly 
positive slopes~\cite{FredricksonJCP2019}.
Since tie lines with exclusively positive 
slopes are also seen for the overall negatively-charged low-pH Ddx4 IDRs
above, the opposite-signed tie-line slopes in Fig.~\ref{fig:50E_KE4b}(a) 
for homopolymeric and those in 
Fig.~\ref{fig:Ddx4_all_salt_lb_high}(c) and (d)
for heteropolymeric polyelectrolytes suggest a role
of sequence heterogeneity in determining whether charged polymers tend to 
exclude or coalesce with salt ions. 
However, the precise origins of variation in tie-line slope remains to
be ascertained. One idea is that the non-zero tie-line slopes arise
from chain connectivity of polymers. If the polymers were
not connected and behave like a collections of monomers, the salt
concentrations in the dilute and condensed phases would simply follow
that of the polymer leading to positive slope \cite{Zhang2018}. However chain
connectivity can change the slope from positive to negative. 

The nature of tie-line slopes has also received considerable attention in the salt-polymer phase diagrams observed
during complex coacervation of symmetric polyelectrolytes\cite{PerryandSing2015,Lytle2016,SingMacro2017,Zhang2018,MuthuJCP2018,TirrellMacro2018,PerrySing2019}. 
Insights gleaned from these studies can yield clues to tie-line slope differences observed in our analysis. A recent theory \cite{Zhang2018} 
based on the concept of chain connectivity predicts a salt-concentration-dependent 
change of sign of tie-line slope, exhibiting a behavior similar to that in Fig.~\ref{fig:Ddx4_all_salt_lb_high}(a) 
and (b). Although in this case of coacervation the slope changes from positive to negative with addition of salt, 
opposite to the case of heteropolymers described here. Another idea is that tie-line slope is determined by a 
competition between electrostatic interactions among polymers and configurational entropy of the salt 
ions,\cite{MuthuJCP2018} whereby the magnitude of electrostatic interactions in the condensed phase 
are enhanced by reduced salt because of less screening but any difference in concentration in salt ions 
between the dilute and condensed phases is entropically unfavorable. It is intuitive that both of these 
proposed mechanisms -- conjectured in modeling coacervation -- would be affected by the charge 
pattern of the polymers, but the manner in which the proposed mechanisms are modulated by sequence 
heterogeneity remains to be investigated.

\subsection{rG-RPA rationalizes sequence-dependent LLPS of Ddx4 IDRs}
Simple RPA theory and an extended RPA+FH theory with an augmented 
Flory-Huggins (FH) mean-field account of non-electrostatic interactions was
utilized to rationalize~\cite{Lin2016, Lin2017a, Brady2017} 
experimental data on sequence- and salt-dependent LLPS of 
Ddx4 IDRs~\cite{Nott2015, Brady2017}. 
Because RPA accounts only for electrostatic interactions and a 
sequence-specific analytical treatment of other interactions is currently
lacking, FH was used to provide an approximate account of non-electrostatic 
interactions.  These interactions can include hydrophobicity, 
hydrogen bonding, and especially cation-$\pi$ and $\pi$-$\pi$ interactions
because $\pi$-related interactions play prominent roles in LLPS of
biomolecular condensates~\cite{Vernon2018}. To gain further insight into the 
semi-quantitative picture emerged from these earlier 
studies~\cite{Lin2016, Lin2017a, Brady2017} and to assess the generality of our 
rG-RPA theory, here we apply an augmented rG-RPA to the LLPS of the same 
Ddx4$^{\rm N1}$ and Ddx4$^{\rm N1}$CS sequences by adding to the rG-RPA 
free energy in Eq.~\ref{f_eq} an FH interaction term
$-\chi \phi_m^2$, where $\chi =  \Delta H (\lb/l) - \Delta S$ contains
both enthalpic and entropic components, and refer to the resulting
formulation as rG-RPA+FH. 

To compare with experimental data~\cite{Brady2017}, we use this theory
to compute the phase diagrams of $\DdxN{}$ and $\DdxNCS{}$ at pH 6.5 
with 100 and 300ml NaCl, which correspond, respectively, 
to $\phi_s = 0.0018$ and 0.0054.
Naturally, pH-dependent behaviors can also be obtained by 
the same FH term together with Eq.~\ref{f_eq} and
Eq.~\ref{eq:ch_pH-dept} for rG-RPA free 
energy; but here we do not pursue a pH-dependent rG-RPA+FH analysis of 
$\DdxN{}$ and $\DdxNCS{}$ LLPS because no corresponding experimental data 
is currently available for comparison. 

\begin{figure}[h!]
   \includegraphics[width=\columnwidth]{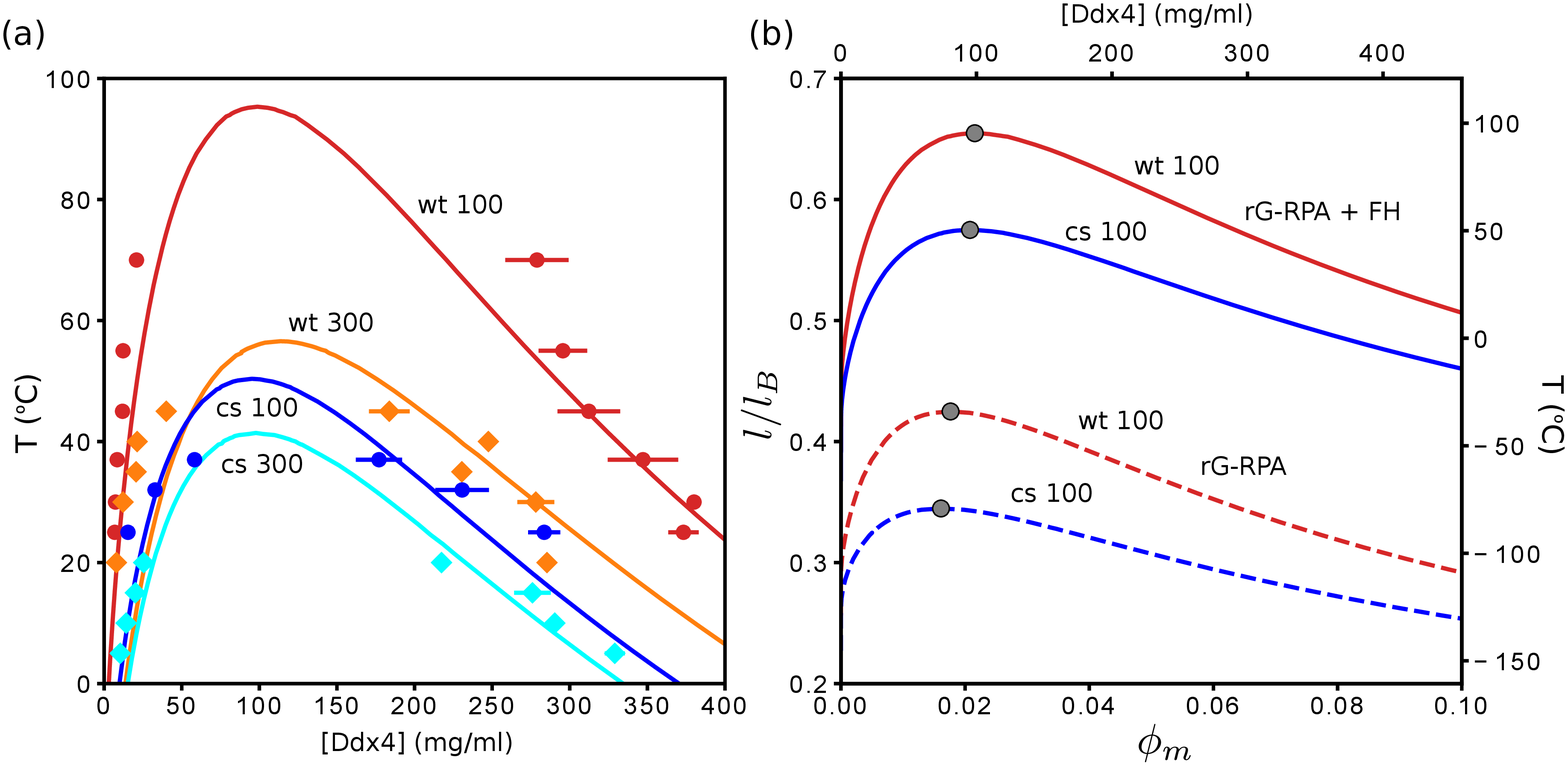} 
   \caption{Comparing rG-RPA+FH results with experimental data on Ddx4 IDRs. 
(a) Experimental data of $\DdxN{pH6.5}$ (wt) and $\DdxNCS{pH6.5}$ (cs) 
(chain length $N=241$ for both sequences)
in aqueous solutions with 100 and 300mM NaCl 
(from Ref.~\citen{Brady2017}; color symbols) are 
fitted, respectively, to rG-RPA+FH theory with $\phi_s=0.0018$ and 0.0054
(continuous curves with the same color). For simplicity,
the salt concentrations in the dilute and condensed Ddx4 phases are taken
to be identical in this calculation. This is a reasonable
approximation because the salt-$\DdxN{}$ binary phase diagrams in 
Fig.~\ref{fig:Ddx4_all_salt_lb_high} indicate that the difference 
in salt concentration between the two phase is less than 15\% for 
$\phi_s < 0.01$. The fits yield an FH interaction parameter
$\chi = 0.5(\lb/l)$ which is equivalent to an enthalpy 
$\Delta H = -0.56$kcal/mol favorable to polymer-polymer attraction.
Model temperatures and model polymer volume fractions
are converted, respectively, to $^\circ$C and mg/ml by 
a procedure similar to that in Ref.~\citen{Lin2016} with
an appropriately chosen model Kuhn length $l$ that is quite similar to
(though not identical with) the C$_\alpha$--C$_\alpha$ virtual bond
length of polypeptides.
(b) Phase diagrams of the two sequences with and without the augmented 
FH interaction. Without the FH term (i.e., $\chi=0$), the critical 
temperatures of 
both $\DdxN{pH6.5}$ and $\DdxNCS{pH6.5}$ at 100mM NaCl are below 0$^{\circ}$C. 
The two $\chi=0$ systems may be interpreted 
as corresponding to sequences with reduced favorable non-electrostatic
interactions~\cite{Brady2017,Vernon2018}. See the main text for further
discussion.}
   \label{fig:jacobfit}
\end{figure}

Our detailed rG-RPA study of salt-$\DdxN{}$ and salt-$\DdxNCS{}$ binary phase 
diagrams in Fig.~\ref{fig:Ddx4_all_salt_lb_high} 
and Figs.~\ref{fig:Ddx4_N1_pH7_salt}--\ref{fig:Ddx4_N1_CS_pH1_salt} indicates 
that the difference between dilute- and condensed-phase salt
concentrations is less than 15\% for $\phi_s < 0.01$. Assuming that
this trend is not much affected by non-electrostatic interactions,
here we make the simplifying assumption that salt concentration is constant 
when determining the rG-RPA+FH phase diagrams. Fig.~\ref{fig:jacobfit}(a) shows
the resulting rG-RPA+FH theory with $\chi = 0.5(\lb/l)$ fits reasonably well with all four available experimental 
phase diagrams.

As control, phase diagrams are also computed without the augmented FH 
term (i.e., $\chi =0$).  These phase diagrams are shown as dashed lines in Fig.~\ref{fig:jacobfit}(b). 
Without the $\chi$ term, the critical temperatures of $\DdxN{}$ and $\DdxNCS{}$ with [NaCl] = 100mM are both 
predicted to be below 0$^{\circ}$C (Fig.~\ref{fig:jacobfit}(b)). This theoretical 
trend is consistent with the experimental observation that phenylalanine
to alanine (F-to-A) and arginine to lysine (R-to-K)
mutants of $\DdxN{}$ do not undergo LLPS at physiologically relevant
temperatures~\cite{Nott2015, Brady2017,Vernon2018}. These mutations (F-to-A and
R-to-K) are expected to significantly reduce $\pi$-related interactions~\cite{Vernon2018} and therefore 
correspond to having a weaker FH term (i.e. $\chi$).

One aforementioned experimentally observed feature that cannot be captured 
by the present rG-RPA+FH theory is that in the absence of salt, $\DdxN{}$ at
pH 6.5 does not phase separate at room temperature, but rG-RPA+FH with
$\chi = 0.5(\lb/l)$ predicts phase separation under the same conditions.
There can be multiple reasons for this mismatch between theory and
experiment, a likely one of which is that the mean-field treatment
of non-electrostatic interactions does not take into possible coupling
(cooperative effects) between sequence-specific electrostatic and 
non-electrostatic interactions such as $\pi$-related interactions and 
hydrogen bonding that can be enhanced by proximate electrostatic
attraction.

\section{Conclusions}
In summary, we have developed a formalism for salt-, pH-, and 
sequence-dependent LLPS by combining RPA and Kuhn-length renormalization. 
The trends predicted by the resulting rG-RPA theory are consistent with 
established theoretical and experimental results. Importantly,
unlike more limited previous analytical approaches, rG-RPA
is generally applicable to both polyelectrolytes and neutral/near-neutral 
polyampholytes. In addition to providing physical rationalizations for 
experimental data on the pH-dependent LLPS of IP5 peptides and sequence 
and salt dependence of LLPS of Ddx4 IDRs, our theory offers several
intriguing predictions of electrostatics-driven LLPS properties that should 
inspire further theoretical studies and experimental evaluations.
One such observation is that in a salt-heteropolymer system, it is possible
for the slope of the tie lines to shift from negative to positive  
by increasing salt. Although tie lines with exclusively positive or exclusively
negative slopes were predicted for uniformly charged polyelectrolytes 
and diblock polyampholytes~\cite{FredricksonJCP2019, Moreira2001, 
Zhang2016, Lytle2016, Lytle2017, Shen2018}, 
a salt-dependent change in
the sign of tie-line slope for a single species of heteropolymer--- specifically from 
negative to positive with increasing salt---is a notable prediction. 
In future studies, it would be interesting to 
explore how this property might have emerged
from the intuitively higher degree of sequence 
heterogeneity of the Ddx4$^{\rm N1}$ IDR vis-\`a-vis that of simple 
diblock or few-block polyampholytes. 
In general, the interplay between sequence heterogeneity and a proposed 
chain connectivity effect\cite{Zhang2018} as well as a proposed 
screening-configurational entropy competition effect\cite{MuthuJCP2018}
on the salt partitioning slope between dilute 
and condensed phases remains to be elucidated.
Another observation of our work 
is that inverse S-shape coexistence curves can arise
from sequence heterogeneity not only for polyampholytes~\cite{Lin2016, Lin2017a,
Lin2017b} but also for polyelectrolytes. 
As emphasized recently~\cite{Das2018b},
an inverse S-shape coexistence curve allows for a less concentrated
condensed phase, which can be of biophysical relevance because it would
enable a condensate with higher permeability~\cite{Wei2017}.

Because rG-RPA is an analytical theory, pertinent numerical computations 
are much more efficient than field-theory or explicit-chain
simulations. Thus, in view of the above advances and despite 
its approximate nature, rG-RPA should be useful as a 
high-throughput tool for assessing sequence-dependent LLPS properties
in developing basic biophysical 
understanding and in practical applications such as design of 
new heteropolymeric materials.

Future development of LLPS theory should address a number of physical
properties not tackled by our current theories.
These include, but not necessarily limited to:
(i) Sequence-dependent effects of non-electrostatic interactions, which 
is neglected 
in rG-RPA+FH. (ii) Counterion condensation~\cite{Manning1979, Orkoulas2003,
MuthuJCP2004,Shen2018}. (iii) Dependence of relative permittivity
(dielectric constant) on polymer density~\cite{Lin2017a,Lin2017c}
and salt~\cite{Orland2012}. 
(iv) A more accurate treatment of conformational
heterogeneity to compute the structure factor. The present approach accounts approximately for 
sequence-dependent end-to-end distance, but it fails to capture
conformational heterogeneities at smaller length scales \cite{Shen2018}. A formalism 
for residue-pair-specific renormalized Kuhn
length~\cite{Sawle2015,GhoshPEStrucFac} should afford improvement in
this regard.
(v) Higher-order density fluctuations beyond the quadratic
fluctuations \cite{Muthu1996} treated by rG-RPA. The rapidly expanding repertoire of 
experimental data on biomolecular condensates is providing impetus 
for theoretical efforts in all these directions.

\section{Acknowledgement}
We thank Alaji Bah, Julie Forman-Kay, and Kevin Shen for helpful discussions. 
This work was supported by
Canadian Insitutes of Health Research grant PJT-155930
and Natural Sciences and Engineering Research Council of Canada
grant RGPIN-2018-04351 to H.S.C., 
National Institutes of Health grant 1R15GM128162-01A1 to K.G.,
and computational resources provided by SciNet of Compute/Calcul Canada.
H.S.C. and K.G. are members of the Protein Folding and Dynamics 
Research Coordination Network funded by National Science Foundation 
grant MCB 1516959. 

\appendix







\section{Derivation of polymer solution free energy}

As described in main text, we consider a neutral solution of $\np$ charged polymers of $N$ monomers (residues) with charge sequence $\ket{\sigma} = [\sigma_1, \sigma_2, ... \sigma^N]^{\rm T}$. Averaged net charge per monomer is defined as $\pc = (\sum_\tau \sigma_\tau)/N$. 
In addition, there are $\ns$ salt ions (co-ions) carrying $\zs$ charges and $\nc$ counterions carrying $\zc$ charges.
Charge neutrality $|\pc| \np + \zs \ns = \zc \nc$ is always preserved. Monomer and ion densities are defined as $\rho_m = \np N/\Omega$, $\rho_s =  \ns/\Omega$, and $\rho_c = \nc/\Omega$, respectively, with 
$\Omega$ being the solution volume.

We label the polymers by $\alpha=1,2,\dots,\np$ and residues in a polymer by
$\tau=1,2,\dots,N$, and denote the spatial coordinate of the $\tau$th monomer
in the $\alpha$th polymer by $\R_{\alpha,\tau}$. Similarly, the small ions are
labeled by $a=1,2,\dots,\ns+\nc$, in which $1\leq a \leq \ns$ are for salt ions
and $\ns\!+\!1\leq a \leq \ns\!+\!\nc$ are for counterions, with the 
coordinate of the $a$th small ion denoted by $\rr_a$. 
The implicit-solvent partition function is then expressed as an integral 
over all solute coordinates divided by factorials that account for the 
indistinguishability of the molecules within each molecular species in 
the solution, viz.,
\begin{equation}
\Z = \frac{1}{\np!\nc!\ns!\nw!}\int 
		\prod_{\alpha=1}^{\np}\prod_{\tau=1}^N d\R_{\alpha,\tau}
		\prod_{a=1}^{\ns+\nc} d\rr_a
		e^{-\T[\R ]-\U[\R,\rr ]},
	\label{eq:ZiniSI}
\end{equation}
where $\nw$ denotes the number of water molecules,
$\T$ accounts for chain connectivity of the polymers, and $\U$ accounts for
interactions among all solute molecules,
$[\R]$ is shorthand for $[\{\R_{\alpha,\tau}\}]$ and
$[\R,\rr]$ is shorthand for $[\{\R_{\alpha,\tau}\},\{\rr_a\}]$.
Connectivity is enforced
by a sum of Gaussian potentials sharing the same Kuhn length $l$,
which is given by
\begin{equation}
\T[\R] = \frac{3}{2l^2}\sum_{\alpha=1}^{\np}\sum_{\tau=1}^{N-1}
\left(\R_{\alpha,\tau+1}-\R_{\alpha,\tau}\right)^2 \; .
	\label{eq:TiniSI}
\end{equation} 
For simplicity, we assume that interactions in $\U$ are all pairwise, in
which case it takes the form
\begin{equation}
\begin{aligned}
\U[\R,\rr] = & \frac{1}{2}\sum_{\alpha,\beta=1}^{\np}\sum_{\tau,\mu=1}^N 
		\UU_{pp}^{\tau\mu}\!\left(\R_{\alpha, \tau}\!-\! \R_{\beta, \mu} \right)  \\
	& + \sum_{\alpha=1}^{\np}\sum_{\tau=1}^N\sum_{a=1}^{\ns+\nc}
		\UU_{ps}^{\tau a} \!\left( \R_{\alpha,\tau}\!-\!\rr_a  \right)   \\
	& + \frac{1}{2}\sum_{a,b=1}^{\ns+\nc} 
			\UU_{ss}^{ab} \!\left( \rr_a\!-\!\rr_b  \right) \; ,
\end{aligned}
	\label{eq:Uall}
\end{equation}
where $\UU_{pp}$, $\UU_{ps}$, and $\UU_{ss}$ are, respectively, 
monomer-monomer, monomer-ion, and ion-ion interaction potentials.
It should be noted that although self-interactions, that is,
the $(\alpha,\tau)=(\beta,\mu)$ terms for monomers and the $a=b$ terms 
for small ions, are included in the above summation to facilitate 
subsequent formal development of a field-theory description, these 
divergent terms will be regularized in the final free energy expression 
and thus have no bearing on the outcome of our theory.
By introducing 
\begin{subequations}
\begin{align}
\rho^\tau_\kk = & \sum_{\alpha=1}^{\np}e^{i \kk \cdot \R_{\alpha,\tau}}
\; ,\label{eq:rho_tau} \\
c^{s}_\kk = & \sum_{a=1}^{\ns}e^{i\kk \cdot \rr_a} \; ,\\
c^{c}_\kk = & \sum_{a=1}^{\nc}e^{i\kk \cdot \rr_{a+\ns}} \; ,
\end{align}
	\label{eq:rho_all}%
\end{subequations}
as the $\kk$-space density operators for the monomers and small ions,
we rewrite Eq.~\ref{eq:Uall} in $\kk$-space as 
\begin{equation}
\U  =  \frac{1}{2\Omega}\sum_\kk
	\Bigg[ \sum_{\tau,\mu=1}^{N}\rho^\tau_\kk \UU_{pp}^{\tau\mu}(\kk) \rho^\mu_{-\kk} 
	 + 2 \sum_{\tau=1}^{N}\sum_{\gamma=s,c} \rho^\tau_\kk \UU_{ps}^{\tau\gamma}(\kk) c^\gamma_{-\kk}
	+ \!\!\sum_{\gamma,\gamma' = s, c} c^\gamma_\kk \UU_{ss}^{\gamma\gamma'}(\kk)c^{\gamma'}_{-\kk}
\Bigg] \; ,
\end{equation}
where $1/\Omega$ is the standard normalization factor for the Fourier 
transformation, and the general form 
$\UU(\kk) = \int d\rr\, \UU(\rr) \exp(-i\kk\!\cdot\!\rr)$ represents the
interaction potentials in $\kk$-space.  As in Eq.~\ref{eq:Uall} for
$\U[\R,\rr]$, the superscripts of $\UU(\kk)$ are labels for monomers and ions, 
and the subscripts specify the interaction type.  We further define
interaction matrices $\hat{\UU}(\kk)$'s by equating the matrix elements 
$[\hat{\UU}(\kk)]_{\tau\mu}$ with $\UU^{\tau\mu}(\kk)$ 
for $\UU_{pp}$, $\UU_{ps}$, and $\UU_{ss}$. We also define the 
density operator vectors $\ket{\rho_\kk}$ and $\ket{c_\kk}$ such that 
$(\ket{\rho_\kk})_\tau = \rho^\tau_\kk$ and $\ket{c_\kk} = [ c_\kk^s , c_\kk^c  ]^{\rm T}$. 
$\U$ can then be expressed in matrix representation as
\begin{equation}
\U = \frac{1}{2\Omega}\sum_\kk\Bigg[ 
\bra{\rho_{-\kk}} \hat{\UU}_{pp}(\kk) \ket{\rho_{\kk}} 
	+ 2\bra{\rho_{-\kk}}\hat{\UU}_{ps}(\kk) \ket{c_{\kk}}
	+ \bra{c_{-\kk}} \hat{\UU}_{ss}(\kk) \ket{c_{\kk}}
\Bigg] \; .
	\label{eq:U_all-interact}
\end{equation}

The present study foucses on solution systems in which $\U_{ss}$ and 
$\U_{ps}$ are purely Coulombic whereas $\U_{pp}$ has both Coulombic 
and pairwise (two-body) excluded-volume repulsion components. Hence
\begin{subequations}
\begin{align}
\hat{\UU}_{ss}(\kk) = & \frac{4\pi \lb}{k^2}\ket{z}\bra{z} \; ,
\label{eq:UUss} \\
\hat{\UU}_{ps}(\kk) = & \frac{4\pi \lb}{k^2}\ket{\sigma}\bra{z} \; , 
\label{eq:UUps} \\
\hat{\UU}_{pp}(\kk) = &  \frac{4\pi \lb}{k^2}\ket{\sigma}\bra{\sigma} 
+ \wtwo\ket{1_N} \bra{1_N} \; , \label{eq:UUpp}
\end{align}
	\label{eq:UU_all}%
\end{subequations}
where $k \equiv |\kk|$, $\lb \equiv e^2/(4\pi\epsilon\kB T)$ is Bjerrum length
($e$ is electronic charge, $\epsilon$ is permittivity,
$\kB$ is Boltzmann constant, $T$ is absolute
temperature). $\bra{z} = {\rm sign}(\pc)[ \zs, -\zc  ]$ is the vector 
representing the charge valencies (number of electronic charges per ion) 
of salt ions and counterions, respectively, $\wtwo>0$ is the strength of
the two-body excluded volume repulsion between monomers, 
and $\ket{1_N}$ is an $N$-dimensional vector in which every component is 1. 
All elements in the excluded volume matrix $\ket{1_N} \bra{1_N}$ take 
unity value because for simplicity all monomers are taken to be of equal size. 
Substituting the potentials given by Eq.~\ref{eq:UU_all} into the $\UU$ 
function in Eq.~\ref{eq:U_all-interact} yields
\begin{equation}
\U = \frac{1}{2\Omega} \sum_{\kk\neq\kzero} \lambda_k 
	| \braket{\sigma}{\rho_{\kk}} + \braket{z}{c_{\kk}}|^2
	+ \frac{1}{2\Omega} \sum_{\kk} \wtwo |\braket{1_N}{\rho_{\kk}}|^2 \; ,
	\label{eq:Uini}
\end{equation}
where $\lambda_k = 4\pi\lb/k^2$ and $| A_\kk|^2  \equiv A_{-\kk} A_\kk$ 
for arbitrary $\kk$-dependent $A_{\kk}$.
The first summation does not need to include $\kk=\kzero$ 
because this term is proportional to the overall net charge of the 
solution and therefore must be zero because of overall electric 
neutrality of the solution.

\subsection{Field theory for polymer solution}

The Hubbard-Stratonovich transformation is then applied to linearize the 
quadratic form $\U$ in Eq.~\ref{eq:Uini} by introducing
conjugate fields $\psi_\kk$ for charge density and $\w_\kk$ for mass density. 
The partition function $\Z$ in Eq.~\ref{eq:ZiniSI} can then be rewritten
in terms of
\begin{equation}
\begin{aligned}
\Z' = & \int\prod_{\alpha=1}^{\np}\prod_{\tau=1}^N d\R_{\alpha,\tau}
		\prod_{a=1}^{\ns+\nc} d\rr_a
		e^{-\T[\R ]-\U[\R,\rr ]} \\
= & \exp\left\{ -\frac{1}{2\Omega}\wtwo|\braket{1_N}{\rho_{\kk=\kzero}}|^2  \right\} 	
	\prod_{\kk\neq\kzero} 
	\int\!\!\frac{d \psi_\kk d\w_\kk}{2\pi\Omega\sqrt{\lambda_k\wtwo}} 
	\exp\left\{  -\frac{1}{2\Omega} \sum_{\kk\neq\kzero}
		\left[ 	\frac{|\psi_\kk|^2}{\lambda_k} + 	\frac{|\w_\kk|^2}{\wtwo} \right]   
	\right\} \\
& \times 
	\int\!\prod_{\alpha=1}^{\np}\prod_{\tau=1}^N d\R_{\alpha,\tau} \!\!
	\prod_{a=1}^{\ns+\nc} d\rr_a 
	\exp\Biggl\{ - \frac{i}{\Omega} \sum_{\kk\neq\kzero}
			\Big[ \big( \braket{\sigma}{\rho_{-\kk}} \!+ \braket{z}{c_{-\kk}} \big) \psi_{\kk} + 
		\braket{1_N}{\rho_{-\kk}} \w_{\kk} \Big] \\
& \hskip 6.5cm 
			\!-\! \T[\{ \R_{\alpha,\tau} \}]   \Biggr\} \; ,
\end{aligned}
	\label{eq:Zprime}
\end{equation}
where $\Z = \Z'/(\np!\nc!\ns!\nw!)$.
The first term in $\Z'$ is merely the $\kk=\kzero$ component of $\U$, 
which by the definition of $\rho_\kk^\tau$ is equal to
\begin{equation}
\Zzero \equiv \exp\left\{ -\frac{1}{2\Omega}\wtwo|\braket{1_N}{\rho_{\kk=\kzero}}|^2  \right\}
	= \exp\left\{ -\frac{\wtwo (N\np)^2}{2\Omega} \right\} \; .
	\label{eq:Z0}
\end{equation}
The remaining terms in $\Z'$ is a field integral of $\psi$ and $\w$. 
The first component (the latter part of the second line in Eq.~\ref{eq:Zprime}) 
is an exponential of the quadratic self-correlations, and the second term 
(the third and fourth lines in Eq.~\ref{eq:Zprime}) 
is a partition function for the 
polymers and the small ions under the influence of $\psi$ and $\w$, 
which we now symbolize as
\begin{equation}
\begin{aligned}
\Qsol[\psi,\w] \equiv &
	\int\prod_{\alpha=1}^{\np}\prod_{\tau=1}^N d\R_{\alpha,\tau} 
	\prod_{a=1}^{\ns+\nc} d\rr_a  \\
&
\times \exp\left\{ - \frac{i}{\Omega} \sum_{\kk\neq\kzero}
			\Big[ \big( \braket{\sigma}{\rho_{-\kk}} \!+ \braket{z}{c_{-\kk}} \big) \psi_{\kk} + 
				\braket{1_N}{\rho_{-\kk}} \w_{\kk} \Big]
			\!-\! \T[\{ \R_{\alpha,\tau} \}]   \right\} \; .
	\label{eq:Qsol}
\end{aligned}
\end{equation}
By the definitions of $c_\kk$ and $\rho^\tau_\kk$ in Eq.~\ref{eq:rho_all},
the exponent in the integrand of $\Qsol$ may be expressed as
\begin{equation}
\begin{aligned}
- \frac{i}{\Omega} \sum_{\kk\neq\kzero} &
			\Big[ \big( \braket{\sigma}{\rho_{-\kk}} \!+ \braket{z}{c_{-\kk}} \big) \psi_{\kk} + 
				\braket{1_N}{\rho_{-\kk}} \w_{\kk} \Big]
			\!-\! \T[\{ \R_{\alpha,\tau} \}] \\
= & - \frac{i}{\Omega} \sum_{\kk\neq\kzero}\psi_\kk\left[
	 (\ket{z})_s\sum_{i=a}^{\ns}e^{-i\kk\cdot\rr_a} +  (\ket{z})_c\sum_{a=\ns+1}^{\ns+\nc}e^{-i\kk\cdot\rr_a} \right] \\
& \quad - 
	\sum_{\alpha=1}^\np\left[
	\frac{3}{2l^2}\sum_{\tau=1}^{N-1} \left(\R_{\alpha,\tau+1}-\R_{\alpha,\tau}\right)^2
	+  \frac{i}{\Omega}
	\sum_{\kk\neq\kzero}\sum_{\tau=1}^N \big( \sigma_\tau\psi_\kk + \w_\kk \big)e^{-i\kk\cdot\R_{\alpha,\tau}}
	\right] \; ,
\end{aligned}
\end{equation}
where $|z\rangle_s={\rm sign}(\pc)\zs$ for salt ions and
$|z\rangle_c=-{\rm sign}(\pc)\zc$ for counterions as defined above.
The coordinates of individual small ions and polymers 
are decoupled in this expression. Thus, the coordinate integrals in 
$\Qsol$ are also decoupled, allowing it to be written as 
\begin{equation}
\Qsol[\psi, \w] = (\QQ_s[\psi])^\ns(\QQ_c[\psi])^\nc (\Qp[\psi,\w])^\np \; ,
	\label{eq:Qsol_sep}
\end{equation}
where the $\ns$, $\nc$, and $\np$ superscripts are powers, with $\QQ_s$ and 
$\QQ_c$ being the single-molecule partition functions for salt ions and 
counterions, respectively; 
$[\psi]$ is shorthand for $[\{\psi_\kk\}]$ and
$[\psi,\w]$ is shorthand for $[\{\psi_\kk\},\{\w_\kk\}]$.
These single-molecule small-ion partition functions are given by
\begin{equation}
\QQ_{s,c}[\psi] = \int d\rr_{s,c} \exp\left\{ -\frac{i(\ket{z})_{s,c} }{\Omega}\sum_{\kk\neq\kzero}
	\psi_\kk e^{-i \kk \cdot \rr_{s,c}}  \right\} \; ,
	\label{eq:Qsc}
\end{equation}
where the expression for $\QQ_s$ or $\QQ_c$ corresponds, respectively,
to choosing the subscript ``$s$'' or $``c$''  for the ``$s,c$'' notation
in the above Eq.~\ref{eq:Qsc}.
The single-polymer partition function $\Qp$ in Eq.~\ref{eq:Qsol_sep} equals
\begin{equation}
\Qp[\psi, \w] = \int\DD[\R] e^{-\Hp[\R; \psi,\w]} \; ,
	\label{eq:Qp}
\end{equation}
where $\DD[\R] \equiv  \prod_{\tau=1}^N d\R_\tau$, 
$[\R; \psi,\w]$ is shorthand for $[\{\R_\tau\},\{\psi_\kk\},\{\w_\kk\}]$,
and 
\begin{equation}
\Hp[\R; \psi,\w] = \frac{3}{2l^2}\sum_{\tau=1}^{N-1} \left(\R_{\tau+1}-\R_{\tau}\right)^2 
		+ \frac{i}{\Omega}\sum_{\kk\neq\kzero} \sum_{\tau=1}^{N} 
				 (\sigma_\tau \psi_\kk + \w_\kk)	e^{-i\kk\cdot\R_{\tau}} \; .
	\label{eq:Hp}
\end{equation}
It should be noted that the small-ion label $a$ and the polymer label $\alpha$
are not needed in the single-molecule partition functions in
Eqs.~\ref{eq:Qsc} and \ref{eq:Qp}. Collecting results from
Eqs.~\ref{eq:Zprime}, \ref{eq:Z0} and \ref{eq:Qsol_sep} yields
the following formula for $\Z'$: 
\begin{equation}
\Z' = \Zzero\int \prod_{\kk\neq\kzero} 
	\frac{d \psi_\kk d\w_\kk}{2\pi\Omega\sqrt{\lambda_k\wtwo}} 
	\exp\left\{  -\frac{1}{2\Omega} \sum_{\kk\neq\kzero}
		\left[ 	\frac{|\psi_\kk|^2}{\lambda_k} + 	\frac{|\w_\kk|^2}{\wtwo} \right]   
	\!+\! \ns\ln\QQ_s \!+\! \nc\ln\QQ_c \!+\! \np\ln\Qp\right\} \; ,
	\label{eq:Zprime_final}
\end{equation}
where $\Zzero$ is provided by Eqs.~\ref{eq:Z0}, 
$\QQ_s$, $\QQ_c$, and $\QQ_p$ are given by
Eqs.~\ref{eq:Qsc}--\ref{eq:Hp}. 

\subsection{Fluctuation expansion of partition function}

To evaluate Eq.~\ref{eq:Zprime_final} analytically, we first derive a 
mean-field solution at $(\psi,\w)=(\psiavg,\wavg)$
in which the mean conjugated fields $\psiavg$ and $\wavg$ satisfy
the extremum condition
$(\delta\Z'/\delta\psi_\kk)=(\delta\Z'/\delta\w_\kk)=0$, which leads to
the equalities
\begin{subequations}
\begin{align}
\frac{\psiavg_\kk}{\Omega\lambda_k}= & 
	\frac{\ns}{\QQ_s}\left(\frac{\delta\QQ_s}{\delta\psi_\kk}\right)_{(\psiavg,\wavg)}
	+ \frac{\nc}{\QQ_c} \left(\frac{\delta\QQ_c}{\delta\psi_\kk}\right)_{(\psiavg, \wavg)}
	+ \frac{\np}{\Qp} \left(\frac{\delta\Qp}{\delta\psi_\kk}\right)_{(\psiavg, \wavg)},  \\
\frac{\wavg_{\kk}}{\Omega \wtwo} = &
	\frac{\np}{\Qp}\left(\frac{\delta\Qp}{\delta\w_\kk} \right)_{(\psiavg,\wavg)}, 
\end{align}
	\label{eq:mft}%
\end{subequations}
where the subscript $(\psiavg,\wavg)$ indicates that the functional
(field) derivatives are evaluated at the to-be-solved mean conjugated 
fields.  The $\psi$ and $\w$ field are conjugates, respectively, to charge 
density and mass density. By using Eqs.~\ref{eq:Qsc}--\ref{eq:Hp}
and the fact that the averages 
$\avg{\cdots}_x$ 
over the spatial coordinates of the given molecular species 
($x=p$, $s$, or $c$) 
of $\kk$-space density operators 
in Eq.~\ref{eq:rho_all}
are given by $\langle\rho^\tau_\kk\rangle_p=\np 
\langle e^{i \kk \cdot \R_{\tau}}\rangle_p$,
$\langle c^{s}_\kk\rangle_s = {\ns}\langle 
e^{i\kk \cdot \rr_s}\rangle_s$, and $\langle c^{c}_\kk\rangle_c = 
{\nc}\langle e^{i\kk \cdot \rr_c}\rangle_c$ 
because of the decoupling stated above by
Eq.~\ref{eq:Qsol_sep},
the first-order derivatives in Eq.~\ref{eq:mft} are given by
\begin{subequations}
\begin{align}
\frac{n_{s,c}}{\QQ_{s,c}}\frac{\delta\QQ_{s,c}}{\delta\psi_\kk} = &
	 -\frac{i(\ket{z})_{s,c}n_{s,c}}{\Omega} 
\Bigl(\avg{ e^{-i\kk\cdot\rr_{s,c}} }_{s,c}\Bigr)_{(\psi,\w)} 
	 = -\frac{i(\ket{z})_{s,c}}{\Omega}  
\Bigl(\avg{ c^{s,c}_{-\kk} }_{s,c}\Bigr)_{(\psi,\w)} \; , \\
\frac{\np}{\Qp}\frac{\delta\Qp}{\delta\psi_\kk} = &
	 -\frac{i\np}{\Omega} 
\Biggl(\avg{ \sum_{\tau=1}^N\sigma_\tau e^{-i\kk\cdot\R_\tau} }_p\; \Biggr)_{(\psi,\w)} 
	 \!\! =  -\frac{i}{\Omega} \sum_{\tau=1}^N \sigma_\tau 
\Bigl(\avg{  \rho_{-\kk}^\tau }_p\Bigr)_{(\psi,\w)}  \; , \\
\frac{\np}{\Qp}\frac{\delta\Qp}{\delta\w_\kk} = &
	 -\frac{i\np}{\Omega} 
\Biggl(\avg{ \sum_{\tau=1}^N e^{-i\kk\cdot\R_\tau} }_p\; \Biggr)_{(\psi,\w)}
	 \!\! =  -\frac{i}{\Omega} \sum_{\tau=1}^N 
\Bigl(\avg{  \rho_{-\kk}^\tau }_p\; \Bigr)_{(\psi,\w)} \; ,
\end{align}
	\label{eq:dQ_dpsi_dw}%
\end{subequations}
where $(\avg{\cdots}_x)_{(\psi,\w)}$ denotes averaging over the spatial 
coordinates of the given molecular species 
evaluated for any given conjugate field $\psi, \w$. 
With Eq.~\ref{eq:dQ_dpsi_dw}, the relations in Eq.~\ref{eq:mft} for 
the mean conjugate fields become 
\begin{equation}
\psiavg_\kk = -i\lambda_k
\Bigl(\avg{\big[ \braket{\sigma}{\rho_{-\kk}}+  
\braket{z}{c_{-\kk}}\big]}_{s,c,p}\Bigr)_{(\psiavg,\wavg)} \; , 
  \quad \quad
\wavg_\kk =  -i\wtwo
\Bigl(\avg{\big[  \braket{1_N}{\rho_{-\kk}}  \big]}_p\Bigr)_{(\psiavg,\wavg)}
\; ,
	\label{eq:mft_general}
\end{equation}
which can now be solved self-consistently to determine 
$\psiavg_\kk$ and $\wavg_\kk$.

We proceed to obtain an approximate solution by assuming that 
within regions where the system exists as a single phase, the mass density is 
rather homogeneous. In that case, the $\kk\ne \kzero$ components
of the density operators $\rho^\tau_\kk$, $c^{s}_\kk$, and $c^{c}_\kk$
in Eq.~\ref{eq:rho_all} are small (approximately zero). It then follows
from Eq.~\ref{eq:mft_general} that
\begin{equation}
\psiavg_\kk \approx \wavg_\kk \approx 0 \; , \quad \forall \kk\neq\kzero \; .
	\label{eq:mft_homo}
\end{equation}
These considerations imply that the following approximate relations
hold for the averaged densities on the right-hand 
side of Eq.~\ref{eq:dQ_dpsi_dw}: 
\begin{equation}
\avg{ c^{s,c}_{-\kk} }_{\approx 0} \approx n_{s,c}\delta_{\kk,\kzero} 
\; , \quad\quad
\sum_{\tau=1}^N \sigma_\tau \avg{  \rho_{-\kk}^\tau }_{\approx 0} \approx \pc \np N\delta_{\kk,\kzero} 
\; , \quad \quad
\sum_{\tau=1}^N \avg{  \rho_{-\kk}^\tau }_{\approx 0} \approx \np N\delta_{\kk,\kzero}
\; ,
	\label{eq:rho_homo}
\end{equation}
where the ``$\approx 0$'' subscript in $\langle\cdots\rangle_{\approx 0}$ 
signifies that the given average over the $s$, $c$, or $p$ spatial coordinates 
is evaluated at the conjuagate fields in Eq.~\ref{eq:mft_homo} for 
approximate homogeneous densities.  Now, to arrive at a definite
approximate description, we expand the logarithmic small-ion partition 
functions around $\psi_{\kk\ne \kzero}=0$ up to $O(\delta\psi^2)$. 
Utilizing the expressions for the averaged densities in Eq.~\ref{eq:rho_homo}
and replacing the conjugate field $\psiavg_{\kk\ne \kzero}\approx 0$ 
(Eq.~\ref{eq:mft_homo}) at which the averages are evaluated
by $\psi_{\kk\ne \kzero}=0$, we obtain
\begin{equation}
\begin{aligned}
\ln \QQ_{s,c}[\psi] \approx & \ln \QQ_{s,c}[\psi_{\kk\ne \kzero}=0]  
	+ \sum_{\kk\neq\kzero} \left(\frac{\delta \ln \QQ_{s,c}}{\delta\psi_\kk}\right)_0 \delta\psi_\kk
	+ \frac{1}{2}\sum_{\kk,\kk'\neq\kzero} 
	\left( \frac{\delta^2 \ln \QQ_{s,c}}{\delta\psi_\kk\delta\psi_{\kk'}}\right)_0
		\delta\psi_\kk\delta\psi_{\kk'} \\
= & \ln\Omega - \frac{i|z\rangle_{s,c}}{\Omega} \sum_{\kk\neq\kzero} \avg{e^{-i\kk\cdot\rr_{s,c}}}_0\delta\psi_\kk \\
& \qquad 
	-\frac{z_{s,c}^2}{2\Omega^2} 
	 \sum_{\kk,\kk'\neq\kzero}\left[
	 \avg{e^{-i(\kk+\kk')\cdot\rr_{s,c}}}_0 
	 -\Bigl\langle e^{-i\kk\cdot\rr_{s,c}}\Bigr\rangle_0 \avg{e^{-i\kk'\cdot\rr_{s,c}}}_0  
	 \right] \delta\psi_\kk\delta\psi_{\kk'}  \\
= & \ln\Omega-\frac{z_{s,c}^2}{2\Omega^2} \sum_{\kk\neq\kzero}|\psi_\kk|^2,
\end{aligned}
	\label{eq:lnQsc_exp}
\end{equation}
where the ``0'' subscript in $(\cdots)_0$
indicates that the derivatives are evaluated at $\psi_{\kk\ne \kzero}=0$.
Similarly, replacing the ``$\approx 0$'' subscripts in Eq.~\ref{eq:rho_homo},
here the ``0'' subscript in $\langle\cdots\rangle_0$ 
indicates that the average is evaluated at $\psi_{\kk\ne \kzero}=0$. In the
last line of the above Eq.~\ref{eq:lnQsc_exp}, the
expansion variable $\delta\psi_\kk$ is written as $\psi_\kk$ for every
term in the $\sum_{\kk\ne \kzero}$ summation because the expansion is 
around $\psi_{\kk\ne \kzero}=0$. 
Substituting Eq.~\ref{eq:lnQsc_exp} for the $\ln \QQ_s$ and $\ln\QQ_c$ into
Eq.~\ref{eq:Zprime_final} yields
\begin{equation}
\Z' \approx  \Zzero\int \prod_{\kk\neq\kzero} 
	\frac{d \psi_\kk d\w_\kk}{2\pi\Omega\sqrt{\lambda_k\wtwo}} 
	\exp\left\{  -\frac{1}{2\Omega} \sum_{\kk\neq\kzero}
		\left[ |\psi_\kk|^2\left(\frac{1}{\lambda_k} \!+\! \zs^2\rho_s \!+\! \zc^2\rho_c \right)
			\!+\! \frac{|\w_\kk|^2}{\wtwo} \right] \!+\! \np\ln\Qp 
+ C
\right\} \; ,
	\label{eq:Zprime_approx}
\end{equation}
where $C=(\ns+\nc)\ln\Omega$ will be dropped in subsequent consideration 
because it has no effect on the relative free energies of different 
configurational states. Let the exponent in Eq.~\ref{eq:Zprime_approx} 
without $C$
be denoted as $-\Eng$, then $\Eng$ may be seen as a Hamiltonian of a 
polymer system:
\begin{equation}
\Eng[\psi, \w] =  \frac{1}{2\Omega}\sum_{\kk\neq\kzero}\left[ 
				\vcol_k\psi_{-\kk}\psi_{\kk} 
				+ \frac{\w_{-\kk}\w_{\kk}}{\wtwo} \right]  - 
\np \ln \Qp[\psi,\w] \; ,
	\label{eq:Zp_Ham}
\end{equation}
where
\begin{equation}
\frac{1}{\vcol_k} = \frac{1}{1/\lambda_k + \zs^2\rho_s+\zc^2\rho_c} \equiv \frac{4\pi\lb}{k^2 + \kappa^2} 
	\label{eq:nu_def}
\end{equation}
is merely a Fourier-transformed Coulomb potential with screening 
length $1/\kappa = [4\pi\lb(\zs^2\rho_s+\zc^2\rho_c)]^{-1/2}$. 
We may now express $\Z'$ as a product of three components, viz.,
\begin{equation}
\Z' = \Zzero\Zion\Zp' \; ,
\end{equation}
where $\Zzero$ is defined in Eq.~\ref{eq:Z0}, 
\begin{equation}
\Zion =\prod_{\kk\neq\kzero} \frac{1}{\sqrt{\vcol_k\lambda_k}} 
	= \prod_{\kk\neq\kzero} \left[ 1 + \frac{\kappa^2}{k^2}\right]^{-\frac{1}{2}} \; ,
\label{eq:Z_factors}
\end{equation}
and 
\begin{equation}
\Zp' = \prod_{\kk\neq\kzero} \int \sqrt{\frac{\vcol_k}{\wtwo}}\frac{d\psi_\kk d\w_\kk}{2\pi\Omega} e^{-\Eng[\psi,\w]} \; .
	\label{eq:Zprime_int}
\end{equation}
Accordingly, the complete partition function $\Z = \Z'/(\ns!\nc!\np!\nw!)$ 
provides free energy of the system in units $\kB T$ per volume $l^3$:
\begin{equation}
f = -\frac{l^3}{\Omega} \ln\Z
	= -s + \fion + \fpoly + \fzero \; ,
\end{equation}
where
\begin{align}
-s = & \frac{l^3}{\Omega}\ln(\ns!\nc!\np!\nw!) \; ,\\
\fzero = &  -\frac{l^3}{\Omega}\ln\Zzero  = \frac{\wtwo l^3(\np N)^2}{2\Omega^2}
	=  \frac{l^3}{2}\wtwo\rho_m^2 \; ,
		\label{eq:fzero_final} \\
\fion = & -\frac{l^3}{\Omega}\ln\Zion 
	= \frac{l^3}{2}\sum_{\kk\neq\kzero}\ln\left[ 1 + \frac{\kappa^2}{k^2}\right]
	= -\frac{(\kappa l)^3}{12\pi} + I_0 \; ,
	\label{eq:-lnZs} \\
\fpoly = & -\frac{l^3}{\Omega}\ln\Zp' \; .
\end{align}

\subsection{Small-ion free energy}

The first term of $\fion$ in Eq.~\ref{eq:-lnZs} is the 
standard Debye screening energy.
The second term of $\fion$, $I_0 = l^3\kappa^2 k_{\rm max}$, is formally 
divergent ($k_{\rm max}$ is the maximum $k$ value of the system, corresponding
to the smallest length scale in coordinate space; $I_0\to\infty$ as $
k_{\rm max}\to\infty$) but since it is linearly proportional to 
$\ns$ and $\nc$ (through its dependence on $\kappa^2$, see above), this
formally divergent term is irrelevant to the relative free energies 
of different configurational states of the system~\cite{Muthu1996}. 
As in most analyses, the $\kk$-summation is performed here by replacing
it with a continuous integral over $\kk$-space:
\begin{equation}
\frac{1}{\Omega}\sum_{\kk\neq\kzero} \to \int\frac{d^3 k}{(2\pi)^3} \; .
\label{eq:dk}
\end{equation}
To make our model physically more realistic, however,
we follow Muthukumar~\cite{Muthu2002, Lee2009} who treated
the charge of each small ion as distributed over a finite volume with 
a characteristic length scale comparable to the bare Kuhn length $l$ 
of the polymers. In this treatment, the point-charge expression for
$\fion$ in Eq.~\ref{eq:-lnZs} is replaced by
\begin{equation}
\fion = -\frac{1}{4\pi}\left[ \ln(1+\kappa l) - \kappa l + \frac{1}{2}(\kappa l)^2 \right] \; ,
	\label{eq:fion_final}
\end{equation}
which reduces to $-(\kappa l)^3/(12\pi)$ in Eq.~\ref{eq:-lnZs}, 
as it should, in the limit of $\kappa l \to 0$. In this regard, 
Eq.~\ref{eq:fion_final}---which is used for all rG-RPA and fG-RPA 
applications in the present work---may be viewed as a regularized, 
more physical version of Eq.~\ref{eq:-lnZs}. 

\subsection{Polymer free energy}

We now proceed to derive an approximate, tractable analytical expression for
$\Zp'$ in Eq.~\ref{eq:Zp_H} in the main text and Eq.~\ref{eq:Zprime_int}
by expanding $\ln\Qp$ (defined in Eqs.~\ref{eq:Qp} and~\ref{eq:Hp}) around 
$\psi_{\kk\ne \kzero}=\w_{\kk\ne \kzero}=0$, viz.,
\begin{equation}
\begin{aligned}
\ln\Qp[\psi,\w] = &\ln\Qp[\psi_{\kk\ne \kzero}\!=\!\w_{\kk\ne \kzero}\!=\!0]
\\ &
			+ \sum_{\kk\neq\kzero}\sum_{\tau=1}^N 
			\left(\frac{\delta\ln \Qp}{\delta\ffc^\tau_\kk}\right)_0\ffc^\tau_\kk 
			+ \frac{1}{2}\sum_{\kk,\kk'\neq\kzero}
			\sum_{\tau,\mu=1}^N\left(
			\frac{\delta^2 \ln\Qp}{\delta\ffc^\tau_\kk\delta\ffc^\mu_{\kk'}} 
			\right)_0
			\ffc^\tau_\kk \ffc^\mu_{\kk'} + O(\ffc^3) \\
= & \ln\Omega+\frac{3(N\!-\!1)}{2}\ln\left( \frac{2\pi l^2}{3}\right) 
	 - \frac{1}{2\Omega^2}\sum_{\kk\neq\kzero} \sum_{\tau,\mu=1}^N
		\avg{e^{-i\kk\cdot(\R_\tau-\R_\mu)}}_0\ffc_\kk^\tau\ffc_{-\kk}^\mu + O(\ffc^3) \; ,
\end{aligned}
	\label{eq:lnQp_Taylor}
\end{equation}
where $\ffc^\tau_\kk = \sigma_\tau\psi_\kk + \w_\kk$ and the first
term in the second line vanishes because of Eq.~\ref{eq:rho_homo}. As 
in Eq.~\ref{eq:lnQsc_exp}, the first two constant terms in the last line
of the above equation have no effect on the relative energies of different
configurations of the system and therefore will be discarded for our
present purpose. 
The third term in the last line of Eq.~\ref{eq:lnQp_Taylor} is the intrachain 
monomer-monomer correlation function evaluated at 
$\psi_{\kk\ne \kzero}=\w_{\kk\ne \kzero}=0$. This correlation function is equal 
to that of a Gaussian chain. However, in the presence of intra- and
interchain interactions, a Gaussian-chain description of the polymer
chains in our system is unsatisfactory, as has been demonstrated
by theoretical and experimental 
studies~\cite{Dobrynin2004,Dobrynin2005,DasPappu2013,Dignon2019}
showing that polymers with different net charges and 
heteropolymers with different charge sequences---even when they have
the same net charge---can have dramatically 
different conformational characteristics.  Intuitively, this sequence-dependent 
conformational heterogeneity should apply not only to the case when a 
polymer chain is isolated but also to situations in which polymer chains 
are in semidilute solutions. To account for this fundamental property
in the monomer-monomer correlation function,
we need to include nonzero $\psi_{\kk\ne \kzero}$ and 
$\w_{\kk\ne \kzero}$ fluctuations that arise from 
the higher-order terms in Eq.~\ref{eq:lnQp_Taylor}.
Accordingly, based on a rationale similar to that advanced
in Refs.~\citen{Muthu1996,Shen2017,Shen2018}, we replace the
monomer-monomer correlation function in Eq.~\ref{eq:lnQp_Taylor}
by a correlation function involving {\em arbitrary} fields. This
development leads to 
\begin{equation}
\ln\Qp[\psi, \w] \simeq -\frac{N}{2\Omega^2}\sum_{\kk\neq\kzero}\left[ 
	\Gss_\kk \psi_\kk \psi_{-\kk} +
	\Gdd_\kk \w_\kk \w_{-\kk} + 
	2\Gds_\kk \w_\kk \psi_{-\kk} \right] \; ,
	\label{eq:lnQ_to_G}
\end{equation}
where $\Gss$, $\Gdd$, and $\Gds$ are structure factors of mass and 
charge densities,
\begin{subequations}
\begin{align}
\Gss_\kk = & \frac{1}{N}\sum_{\tau,\mu=1}^N \sigma_\tau \sigma_\mu 
\Bigl(\avg{e^{i\kk\cdot(\R_\tau-\R_\mu)}}_{p}\Bigr)_{(\psi,\w)} \; ,\\
\Gdd_\kk = & \frac{1}{N}\sum_{\tau,\mu=1}^N 
\Bigl(\avg{e^{i\kk\cdot(\R_\tau-\R_\mu)}}_p\Bigr)_{(\psi,\w)} \; ,\\
\Gds_\kk = & \frac{1}{N}\sum_{\tau,\mu=1}^N \sigma_\tau 
\Bigl(\avg{e^{i\kk\cdot(\R_\tau-\R_\mu)}}_p\Bigr)_{(\psi,\w)} \; .
\end{align}
	\label{eq:structure_factors}%
\end{subequations}
Substituting Eq.~\ref{eq:lnQ_to_G} for $\ln\Qp$ in Eq.~\ref{eq:Zp_Ham}, 
we obtain
\begin{equation}
\begin{aligned}
\Eng[\psi, \w] = &  \frac{1}{2\Omega}\sum_{\kk\neq\kzero}
	\left\langle \psi_{-\kk} \; \w_{-\kk} \right|
\left(
\begin{array}{cc}
\vcol_k + \rho_m \Gss_\kk & \rho_m \Gds_\kk \\
\rho_m \Gds_\kk & \wtwo^{-1} + \rho_m \Gdd_\kk
\end{array}
\right)
\left|
\begin{array}{c}
\psi_\kk \\
\w_\kk
\end{array}
\right\rangle \\
= & \frac {1}{2\Omega}\sum_{\kk\neq\kzero} \bra{\ffv_{-\kk}} \hat{\Delta}_\kk \ket{\ffv_\kk} \; ,
\end{aligned}
	\label{eq:Ham_matrix}
\end{equation}
where 
$\bra{\ffv_{-\kk}}\equiv \langle \psi_{-\kk} \; \w_{-\kk}|$,
$\ket{\ffv_\kk}=(\bra{\ffv_{-\kk}})^{*{\rm T}}$, and $\hat{\Delta}_\kk$ 
is the $2\times 2$ matrix in the above equation. 
Thus, each term in the product given in Eq.~\ref{eq:Zprime_int} 
can now be evaluated as a Gaussian integral to yield
\begin{equation}
\Zp' = \prod_{\kk\ne \kzero} \sqrt{ \frac{\vcol_k}{\wtwo \det \hat{\Delta}_\kk}  } \; .
	\label{eq:Zprime_gauss_int}
\end{equation}
Therefore, by Eqs.~\ref{eq:dk} and \ref{eq:Zprime_gauss_int}, 
the unit free energy is now formally given by
\begin{equation}
\fpoly = -\frac{l^3 \ln\Zp'}{\Omega} = \frac{l^3}{2}\int \frac{d^3 k}{(2\pi)^3}
	\ln\left[
1 + \rho_m\left( \frac{\Gss_\kk}{\vcol_k} + \wtwo\Gdd_\kk \right)
+ \frac{\wtwo}{\vcol_k} \rho_m^2 \left( \Gss_\kk\Gdd_\kk - \Gds_\kk^2 \right)
	\right].
	\label{eq:fpolySI}
\end{equation}
It should be noted, however, that 
the $k\equiv |\kk|\to\infty$ behavior of the integrand in the above 
Eq.~\ref{eq:fpolySI} needs to be regularized. For point particles, 
the $k\to\infty$ limit of the pairwise correlation function is
a Kronecker-$\delta$:
\begin{equation}
\lim_{k\to\infty}\avg{e^{i\kk\cdot(\R_\tau-\R_\mu)}}_p = \delta_{\tau\mu}
\; .
\end{equation}
Thus, by Eq.~\ref{eq:structure_factors},
\begin{subequations}
\begin{align}
\lim_{k\to\infty}\Gss_\kk = & \frac{1}{N}\sum_{\tau=1}^N \sigma_\tau^2 
\; ,  \\
\lim_{k\to\infty}\Gdd_\kk = & 1  \; ,\\
\lim_{k\to\infty}\Gds_\kk = & \pc \; .
\end{align}
	\label{eq:structure_factors_k->infty}%
\end{subequations}
Because $\lim_{k\to\infty}(1/\vcol_k) = \lim_{k\to\infty} 4\pi\lb/k^2$ and 
$\wtwo>0$, 
Eq.~\ref{eq:structure_factors_k->infty} indicates that the integral 
in Eq.~\ref{eq:fpolySI} has an ultraviolet (large-$k$) divergence.
This divergence is physically irrelevant, however, because the integral
can be readily regularized by subtracting the unphysical Coulomb 
self-energy of the charged monomers 
\begin{equation}
f_{\rm self} =  \frac{\rho_m l^3}{2N} \int \frac{d^3k}{(2\pi)^3} \frac{4\pi\lb}{k^2}\sum_{\tau=1}^N \sigma_\tau^2
\end{equation}
that was included merely for formulational convenience in the first place.
In the same vein as the charge smearing for the small ions 
(Eq.~\ref{eq:fion_final}), we also smear the $\delta$-function excluded 
volume repulsion by a Gaussian~\cite{Wang2010,Villet2014}, viz.,
\begin{equation}
\wtwo \to \wtwo(k) = \wtwo e^{-\frac{1}{6}(kl)^2} \; ,
	\label{eq:w2_reg}
\end{equation}
and use $\wtwo(k)$ in the integral of Eq.~\ref{eq:fpolySI} of $\fpoly$ to give
a $\wtwo$-regularized $\fpoly[\wtwo(k)]$.
The regularized $\fpoly$ resulting from these two procedures is then given by
\begin{equation}
\fpoly[\wtwo(k)] - f_{\rm self} \rightarrow \fpoly \; ,
	\label{eq:fpoly_reg}
\end{equation}
where the last arrow signifies that this regularized version of
$\fpoly$ is the one used for our subsequent theoretical development
in the present work.

As discussed above, the present separate treatments for small-ions 
(Eq.~\ref{eq:fion_final}) 
and polymers (Eqs.~\ref{eq:fpolySI} and \ref{eq:fpoly_reg})
are needed in our formulation---which expresses the total partition
function as a product consisting of separate factors for small ions
and polymers (Eq.~\ref{eq:Z_factors})---such that the polymer part of 
the partition function can be used to derive an effective Kuhn length.
Not surprisingly, in the event that the bare chain length $l$ is used 
instead of an effective Kuhn length and  that the volume of small ions 
and the volume of the monomers of the polymers becomes negligible
($\wtwo\rightarrow 0$), the free energy expression reduces to 
that of our simple RPA theory~\cite{Lin2016, Lin2017a}, as can be
readily seen in the following. First,
when the size of the small ions is assumed to be negligible, their
free energy is given by the simple Debye-H\"{u}ckel expression in 
Eq.~\ref{eq:-lnZs} instead of the finite-size expression 
in Eq.~\ref{eq:fion_final}. Second, as $\wtwo\rightarrow 0$, all
terms involving $\wtwo$ in Eq.~\ref{eq:fpolySI} vanish. 
Consequently, the resulting overall electrostatic free energy, denoted
here as $f_{\rm el}$, is given by
\begin{equation} 
f_{\rm el} = \fion^{\rm (Eq.\ref{eq:-lnZs})} + \fpoly^{\rm (Eq.\ref{eq:fpolySI}) }(\wtwo\to0) = 
	\frac{l^3}{2}\int \frac{d^3k}{(2\pi)^3}\left\{  
		\ln\left[ 1 + \frac{\kappa^2}{k^2}\right]  +
	 	\ln\left[ 1 + \rho_m\frac{\xi_\kk}{\nu_k}\right]
	 \right\}.		
\end{equation}
Recalling that $\kappa^2 = 4\pi\lb(\zs^2\rho_s + \zc^2\rho_c)$ and
$1/\nu_k = 4\pi\lb/(k^2+\kappa^2)$ (Eq.~\ref{eq:nu_def}), this quantity
becomes 
\begin{equation} 
\begin{aligned}
f_{\rm el} = & \frac{l^3}{2}\int \frac{d^3k}{(2\pi)^3} \ln
	\left[
	\frac{{\kappa^2+k^2}}{k^2} \times \frac{k^2+\kappa^2 + 4\pi\lb\rho_m\Gss_\kk}{{k^2 + \kappa^2}}
	\right] \\
= & \frac{l^3}{2}\int \frac{d^3k}{(2\pi)^3} \ln
	\left[ 1 + \frac{\kappa^2 + 4\pi\lb\rho_m\Gss_\kk}{k^2}
	\right]  \\
= &  \frac{l^3}{2}\int \frac{d^3k}{(2\pi)^3} \ln\left[ 1 + \frac{4\pi\lb}{k^2}\left( \zs^2 \rho_s + \zc^2 \rho_c^2 + \rho_m\Gss_\kk\right)\right],
\end{aligned}
\end{equation}  
which is exactly the same $f_{\rm el}$ expression in our previous simple 
RPA theory in a formulation that does not consider an explicit excluded-volume
repulsion term and treats small ions and polymers on the same 
footing\cite{Lin2016,Lin2017a}.

\subsection{Effective Gaussian-chain model for two-body correlation function}

The $(\psi,\w)$-dependence of the structure factors $\Gss$, $\Gdd$, and $\Gds$ 
in Eq.~\ref{eq:fpolySI} for $\fpoly$ allows for an 
account of sequence-dependence conformational heterogeneity by
using a Gaussian chain with a renormalized Kuhn length~\cite{Sawle2015} 
$\leff = xl$ (instead of the ``bare'' Kuhn length $l$) to approximate
the polymer partition function $\Qp$ in Eq.~\ref{eq:Qp}. Specifically,
we make the approximation that
\begin{equation}
\Qp \approx \int \DD[\R] e^{-\Hp^{0}[\R]} \; , \quad {\rm where} \quad
\Hp^0[\R] = \frac{3}{2l^2 x}\sum_{\tau=1}^{N-1}\left( \R_{\tau+1}-\R_\tau\right)^2.
	\label{eq:Qp_Xapprox}
\end{equation}
The structure factors $\Gss$, $\Gdd$, and $\Gds$ in 
Eq.~\ref{eq:structure_factors} can then be readily expressed
in terms of the yet-to-be-determined renormalization parameter $x$:
\begin{subequations}
\begin{align}
\Gss_\kk \to & \Gss_k^x = \frac{1}{N}\sum_{\tau,\mu=1}^N \sigma_\tau\sigma_\mu 
	e^{-\frac{1}{6}(kl)^2x|\tau-\mu|} \; ,\\
\Gdd_\kk \to & \Gdd_k^x = \frac{1}{N}\sum_{\tau,\mu=1}^N  e^{-\frac{1}{6}(kl)^2x|\tau-\mu|} \; ,\\
\Gds_\kk \to &  \Gds_k^x = \frac{1}{N}\sum_{\tau,\mu=1}^N \sigma_\tau e^{-\frac{1}{6}(kl)^2x|\tau-\mu|}\; .
\end{align}
	\label{eq:SF_x}
\end{subequations}
\noindent
The renormalization parameter $x=\leff/l$ is determined using 
a sequence-specific 
variational approach introduced by Sawle and Ghosh~\cite{Sawle2015, Muthu1987},
as follows. We first express the Hamiltonian $\Hp[\R]$ in Eq.~\ref{eq:Hp} 
as $\Hp = \Hp^0 + \Hp^1$, where $\Hp^0$ 
(given by Eq.~\ref{eq:Qp_Xapprox}) is the principal term and 
\begin{equation}
\Hp^1[\R;\psi,\w] = \frac{3}{2l^2}\left( 1-\frac{1}{x} \right) \sum_{\tau=1}^{N-1} \left( \R_{\tau+1} - \R_\tau\right)^2 
		+ \frac{i}{\Omega}\sum_{\kk\neq\kzero} \sum_{\tau=1}^N \left( \sigma_\tau \psi_\kk + \w_\kk \right)e^{-i\kk\cdot\R_\tau} \; 
	\label{eq:Hp1}
\end{equation}
is the perturbative term. Then,
for any given physical quantity $A[\R]$, the perturbation 
expansion of its thermodynamic average over polymer configurations 
$\{\R_\tau\}$ and field fluctuations $\ffv=(\psi,\w)$ is given 
by~\cite{DoiEdwardsbook}
\begin{equation}
\begin{aligned}
\langle A[\R] \rangle = & \frac{\avg{e^{-\Hp^1[\R;\ffv]} A[\R]}_{0,\ffv}}{\avg{e^{-\Hp^1[\R;\ffv]}}_{0,\ffv}} \\
	= & \avg{ A[\R] }_{0}
		+ \left[ \avg{ A[\R]  }_{0} \avg{ \Hp^1[\R;\ffv] }_{0,\ffv} - \avg{ A[\R] \, \Hp^1[\R;\ffv] }_{0,\ffv}  \right]  \\
		& \qquad\quad\;\; + \frac{1}{2}\left[ \avg{ A[\R] \left(\Hp^1[\R;\ffv]\right)^2 }_{0,\ffv}
			- \avg{ A[\R] }_0 \avg{ \left(\Hp^1[\R;\ffv]\right)^2 }_{0,\ffv} \right] \\
		& \qquad\quad\;\; + \avg{ A[\R] }_0 \avg{ \Hp^1[\R;\ffv] }_{0,\ffv}^2 
				- \avg{ A[\R] \Hp^1[\R;\ffv] }_{0,\ffv} \avg{ \Hp^1[\R;\ffv] }_{0,\ffv}
				   \\
		& \qquad\quad\;\; + O\left( (\Hp^1)^3 \right) \; ,
	\label{eq:variational}
\end{aligned}
\end{equation}
where the subscripts $0,\ffv$ in $\avg{\cdots}$ signify, respectively,
that the average over $\{\R_\tau\}$'s is weighted by the Hamiltonian
$\Hp^0[\R]$ in Eq.~\ref{eq:Qp_Xapprox} and the average over field 
configurations is weighted by the Hamiltonian $\Eng[\psi,\w]$ 
in Eq.~\ref{eq:Zp_Ham}. (Note that the meaning of the ``0'' subscript 
here is different from that for the averages evaluated 
at $\psi_{\kk\ne \kzero}=\w_{\kk\ne \kzero}=0$ in Eq.~\ref{eq:lnQsc_exp}). 
An $\Hp^0[\R]$ that provides a good description of the thermal 
properties of $A$ may then be obtained by minimizing $\avg{A}-\avg{A}_0$.
This is accomplished by a partial optimization to seek a value of $x=\leff/l$
that would abolish the lowest-order nontrivial $\Hp^1$ contributions 
in Eq.~\ref{eq:variational}.

To obtain a partially optimized $x=\leff/l$ that provides a good
approximation for the monomer-monomer correlation function, $A$ is
chosen to be the squared end-to-end distance of the polymer, 
i.e., $A=\Ree^2 \equiv |\R_N-\R_1|^2$, because $\Ree$ is a simple yet
effective measure of conformational dimensions of 
polymers~\cite{Muthu1996,Sawle2015}. To facilitate this calculation, 
we express $\Hp^1$ in Eq.~\ref{eq:Hp1} as $\Hp^1 = \ChiA + \ChiB$, where
\begin{subequations}
\begin{align}
\ChiA[\R] = & \frac{3}{2l^2}\left( 1-\frac{1}{x} \right) 
\sum_{\tau=1}^{N-1} \left( \R_{\tau+1} - \R_\tau\right)^2 \; , \\
\ChiB[\R;\ffv] = & \frac{i}{\Omega}\sum_{\kk\neq\kzero} 
\sum_{\tau=1}^N \left( \sigma_\tau \psi_\kk + 
\w_\kk \right)e^{-i\kk\cdot\R_\tau} \; ,
\end{align}
\end{subequations}
such that $\ChiA[\R]$ is independent of $\ffv$ and all of $\Hp^1$'s
dependence on $\ffv$ is contained in $\ChiB[\R;\ffv]$. It follows that
the $\ffv$ average is trivial (i.e., it produces a multiplicative factor 
of unity and therefore can be omitted) for any function of $\ChiA[\R]$ 
only. In Eq.~\ref{eq:variational},
the only contributions from terms linear in $\ChiA[\R]$ come from 
the first line on the right-hand side (after the second equality), which equal
\begin{equation}
\avg{ \Ree^2 }_0 \avg{ \ChiA }_0 - \avg{ \Ree^2 \ChiA }_0
	= -l^2 (N-1) x(x-1) \; .
	\label{eq:Chi1_perturb}
\end{equation}
For the $\ChiB$-containing terms in Eq.~\ref{eq:variational}, we
first consider their $\ffv$-averages before applying the $\avg{\cdots}_0$
averaging. For terms linear in $\ChiB$, it is straightforward to see
that
\begin{equation}
\avg{\ChiB}_\ffv = \frac{i}{\Omega}\sum_{\kk\neq\kzero}\sum_{\tau=1}^N
	\Big[ \sigma_\tau\avg{\psi_\kk}_\ffv + \avg{\w_\kk}_\ffv  \Big]e^{-i\kk\cdot\R_\tau} = 0
\end{equation}
because $\avg{\psi_\kk}_\ffv = \avg{\w_\kk}_\ffv = 0$ according to 
the quadratic-field Hamiltonian $\Eng[\psi, \w]$ in Eq.~\ref{eq:Ham_matrix}. 
Thus, $\ChiB$ has zero contribution in the first and third 
lines on the right-hand side of Eq.~\ref{eq:variational}.
In contrast, terms quadratic in $\ChiB[\R]$ are not identical zero,
because
\begin{equation}
\avg{\ChiB^2}_\ffv = -\frac{1}{\Omega^2}\sum_{\kk\neq\kzero}
	\sum_{\tau,\mu=1}^N
	\Big[  \sigma_\tau\sigma_\mu \avg{\psi_{-\kk}\psi_{\kk}}_\ffv
		+ \avg{\w_{-\kk}\w_{\kk}}_\ffv
		+ (\sigma_\tau \!+\! \sigma_\mu) \avg{\psi_{-\kk}\w_{\kk}}_\ffv
	 \Big] e^{-i\kk\cdot(\R_\tau-\R_\mu)} \; ,
\label{eq:chib}
\end{equation}
and here $\avg{\ChiB^2}_\ffv$ is seen as depending on field-field correlation 
functions $\avg{\psi\psi}$, $\avg{\w\w}$, and $\avg{\psi\w}$ averaged 
over $\ffv$. Thus, the $\ChiB^2$ factors in the averages in 
the second line on the right-hand side of Eq.~\ref{eq:variational}
provide the only nonzero contribution through second order in $\Hp^1$.
Following Ref.~\citen{Muthu1996}, we only consider lowest-order nonzero 
contributions from $\ChiA$, and from $\ChiB$, separately, i.e., including only
terms through $O(\ChiA)$ and $O(\ChiB^2)$ as discussed above. 
This approach to the perturbative analysis of Eq.~\ref{eq:variational} may
also be rationalized by an alternate analytical formulation put forth
in Refs.~\citen{Shen2017,Shen2018}.

As shown in Eq.~\ref{eq:Ham_matrix}, the field configuration distribution 
may be approximated by a Gaussian distribution embodied by
the quadratic Hamiltonian $\Eng[\psi,\w]$.
According to perturbation theory~\cite{CardyBook, Lin2017a}, the field-field 
correlation functions in Eq.~\ref{eq:chib} can now be obtained from the 
matrix $\hat{\Delta}_\kk$ in Eq.~\ref{eq:Ham_matrix} via the relationships
\begin{subequations}
\begin{align}
\frac{\avg{\psi_{-\kk}\psi_\kk}}{\Omega} = & 
	\left( \hat{\Delta}_\kk^{-1}\right)_{11} = \frac{\wtwo^{-1}+\rho_m\Gdd_\kk}{\det \hat{\Delta}_\kk} \; , \\
\frac{\avg{\w_{-\kk}\w_\kk}}{\Omega} = & 
	\left( \hat{\Delta}_\kk^{-1}\right)_{22} =   \frac{\vcol_k + \rho_m \Gss_\kk}{\det \hat{\Delta}_\kk} \; , \\
\frac{\avg{\psi_{-\kk}\w_{\kk} }}{\Omega} = \frac{\avg{\psi_{\kk}\w_{-\kk} }}{\Omega} = & 
	\left( \hat{\Delta}_\kk^{-1}\right)_{12} = \left( \hat{\Delta}_\kk^{-1}\right)_{21} 
	= \frac{-\rho_m\Gds_\kk}{\det \hat{\Delta}_\kk} \; .
\end{align}
\end{subequations}
Hence $\avg{\ChiB^2}$ is expressed in terms of $\hat{\Delta}_\kk$ as
\begin{equation}
\begin{aligned}
\avg{\ChiB^2} = & -\frac{1}{\Omega}\sum_{\kk\neq\kzero}\sum_{\tau,\mu=1}^N
	\frac{ \left\langle\sigma_\tau \;1 \right|  \left(
	\begin{array}{cc}
	\wtwo^{-1} + \rho_m\Gdd_\kk & -\rho_m\Gds_\kk \\
	-\rho_m\Gds_\kk & \vcol_k + \rho_m\Gss_\kk 
	\end{array} \right)
	\left|  
	\begin{array}{c}
	\sigma_\mu \\
	1
	\end{array}
	\right\rangle}
	{\det \hat{\Delta}_\kk} 
	e^{-i\kk\cdot(\R_\tau-\R_\mu)} \; .\\
\end{aligned}
	\label{eq:X2_avg_final}
\end{equation}
It should be noted that the excluded volume interaction
$\wtwo$ is not regularized by Eq.~\ref{eq:w2_reg} here because a 
$k$-independent $\wtwo$ is needed to guarantee a real solution
for the renormalization parameter $x$ for arbitrary charge sequence 
$\ket{\sigma}$ (Refs.~\citen{Sawle2015,Firman2018}). Thus, the regularized
form of $\wtwo$ in Eq.~\ref{eq:w2_reg} applies only to the explicit
$v_2$ dependence of $\fpoly$ in Eq.~\ref{eq:fpolySI} but not the implicit 
$v_2$ dependence of $x$ contained in the renormalized form of the 
structure factors $\Gss$, $\Gdd$, and $\Gds$.
Substituting the $x$-dependent correlation functions in Eq.~\ref{eq:SF_x} for
the structure factors in Eq.~\ref{eq:X2_avg_final}, we obtain the nonzero
contribution from $\ChiB$ in the second line of the right-hand side
of Eq.~\ref{eq:variational} as
\begin{equation}
\frac{1}{2}\left[ \langle \Ree^2 \ChiB^2 \rangle_0
			- \langle \Ree^2 \rangle_0 \langle \ChiB^2 \rangle_0 
	 	\right]
	=  \frac{N  l^4 x^2}{18} \int \!\!\frac{d^3 k}{(2\pi)^3}
		 \frac{k^2 \Xee_k^x}{\det \Delta_k^x} \; ,
	\label{eq:Chi2_perturb}
\end{equation}
where 
\begin{equation}
\det \Delta_k^x = \frac{\vcol_k}{\wtwo} + \rho_m\left( \frac{\Gss_k^x}{\wtwo} + \vcol_k \Gdd_k^x \right) 
	+ \rho_m^2\left[ \Gss_k^x \Gdd_k^x - \right(\Gds_k^x\left)^2 \right] 
\; ,
	\label{eq:detDel}
\end{equation}
and 
\begin{equation}
\Xee_k^x \equiv \frac{\LGss_k^x}{\wtwo} + \vcol_k \LGdd_k^x + 
	\rho_m\left( \LGss_k^x \Gdd_k^x + \Gss_k^x\LGdd_k^x - 2\Gds_k^x \LGds_k^x \right) \; .
\end{equation}
Here the renormalized $\LGss$, $\LGdd$, and $\LGds$ are given by
\begin{subequations}
\begin{align}
\LGss_k^x = & \frac{1}{N}\sum_{\tau,\mu=1}^N \sigma_\tau\sigma_\mu|\tau-\mu|^2 e^{-\frac{1}{6}(kl)^2x |\tau-\mu|} \; ,\\
\LGdd_k^x = & \frac{1}{N}\sum_{\tau,\mu=1}^N  |\tau-\mu|^2 e^{-\frac{1}{6}(kl)^2x |\tau-\mu|}\; , \\
\LGds_k^x = & \frac{1}{N}\sum_{\tau,\mu=1}^N \sigma_\tau |\tau-\mu|^2 e^{-\frac{1}{6}(kl)^2x |\tau-\mu|} \; .
\end{align}
\end{subequations}
Finally, by combining Eqs.~\ref{eq:Chi1_perturb} and \ref{eq:Chi2_perturb}, 
we arrive at the variational equation 
\begin{equation}
1-\frac{1}{x} - \frac{N l^2}{18(N-1)} \int \!\!\frac{d^3 k}{(2\pi)^3} \frac{k^2  \Xee_k^x}{\det \Delta_k^x} = 0
	\label{eq:x_sol_finalSI}
\end{equation}
for solving $x$.
In our numerical calculations, we take $\wtwo=4\pi l^3/3$. Inserting 
the solution of $x$ into Eq.~\ref{eq:SF_x} provides
an improved accounting of the conformational heterogeneity
in the free energy; and this improvement is central to the
present rG-RPA theory. 

\subsection{Mixing entropy}

The factorials in Eq.~\ref{eq:ZiniSI} 
arise from the indistinguishability of the molecules belonging to the
same species. Taking logarithm and using Stirling's approximation,
one obtains
\begin{equation}
\begin{aligned}
-\frac{S}{\kB} = & \ln\left( \np!\ns! \nc! n_\w!  \right) \\
\simeq & 
\np \ln \np +\ns\ln \ns + \nc \ln \nc + n_w \ln n_w - \np - \nc - \ns - \nw \; ,
\end{aligned}
	\label{eq:Sini}
\end{equation}
where additive terms of the form $[\ln(2\pi n)]/2$ (where 
$n = \np$, $\ns$, $\nc$, or $\nw$) are omitted because for large $n$, 
their contributions is negligible in comparison to the terms included 
in Eq.~\ref{eq:Sini}.
As in Ref.~\citen{Lin2017a}, here we assume for simplicity that 
the size of a monomer, a small ion, or a water molecule all equals $l^3$.
Assuming further, for simplicity, that the system is incompressible,
i.e., the system volume $\Omega$ is fully occupied by polymers, small
ions, and water, then
\begin{equation}
\frac{1}{\Omega}\left( N \np + \ns + \nc + \nw \right) = 
\rho_m + \rho_s + \rho_c + \rho_w = \frac{1}{l^3} \; .
\end{equation}
Following Flory's notation, volume fractions of polymers and 
salt ion are defined, respectively, as
\begin{equation}
\phi_m = \rho_m l^3 \; , \; \phi_s = \rho_s l^3 \; ,
\end{equation}
and the volume fraction $\phi_c$ of counterions
and volume fraction $\phi_w$ of water are given by
\begin{equation}
\zc \phi_c = \pc \phi_m + \zs \phi_s \; , \; \phi_w = 1-\phi_m-\phi_s-\phi_c
\; .
	\label{eq:phic_phiw}
\end{equation}
Because the last four terms in Eq.~\ref{eq:Sini} are linear
in numbers of molecules, they are irrelevant to phase
separation~\cite{Lin2017a}. Discarding these 
terms results in the mixing entropy
\begin{equation}
-s \equiv -\frac{S l^3}{\kB \Omega}
	= \frac{\phi_m}{N}\ln\phi_m
	+ \phi_s \ln \phi_s
	+ \phi_c \ln\phi_c
	+ \phi_w\ln \phi_w
	\label{eq:s_final}
\end{equation}
given in Eq.~2 of the main text. 


\section{Temperature selection for polymer-salt phase diagrams of 
Ddx4 variants}

Three temperatures, two below and one slightly above the respective
salt-free critical temperature $T_{\rm cr}\propto l/(\lb)_{\rm cr}$
of each of the Ddx4 variants $\DdxNn$, $\DdxNCSn$, $\DdxNl$, and $\DdxNCSl$ 
are selected for the phase diagrams in Figs.~\ref{fig:Ddx4_N1_pH7_salt}, 
\ref{fig:Ddx4_N1_CS_pH7_salt}, \ref{fig:Ddx4_N1_pH1_salt}, and 
\ref{fig:Ddx4_N1_CS_pH1_salt}.
The $l/\lb$ values are selected to compare salt dependence of the 
sequences under temperatures producing similar gaps between 
the dilute- and condensed-phase protein densities at or near $\phi_s=0$ 
for the different sequences. Specifically, for the same part
of the figures ((a), (b), and (c) separately), the $l/(\lb)$'s
are such that dilute-condensed density gaps are similar across 
Figs.~\ref{fig:Ddx4_N1_pH7_salt}--\ref{fig:Ddx4_N1_CS_pH1_salt}.


\bibliography{rgRPA_JCP_Final}

\bibliography{rgRPA_JCP_Final}

\end{document}